\documentclass[twocolumn,aps,prl,superscriptaddress]{revtex4}

\usepackage{amsmath}
\usepackage{graphicx}
\usepackage{MnSymbol}
\usepackage{comment}

\begin{document}

\title{Theory for diffusive and ballistic air leakage and its application to suction cups}

\author{A. Tiwari}
\affiliation{PGI-1, FZ J\"ulich, Germany, EU}
\author{B.N.J. Persson}
\affiliation{PGI-1, FZ J\"ulich, Germany, EU}
\affiliation{www.MultiscaleConsulting.com}

\begin{abstract}
We have developed a theory of air leakage at interfaces between two elastic solids with application
to suction cups in contact with randomly rough surfaces. We present an equation for the airflow in narrow constrictions
which interpolate between the diffusive and ballistic (Knudsen) air-flow limits. To test the theory we performed experiments
using two different suction cups, made from soft polyvinylchloride (PVC), in contact
with sandblasted polymethylmethacrylate (PMMA) plates. We found that
the measured time to detatch (lifetime) of suction cups were in good agreement with theory, 
except for surfaces with the root-mean-square (rms) roughness below
$\approx 1 \ {\rm \mu m}$, where diffusion of plasticizer from the PVC to the PMMA surface caused blockage of critical
constrictions. Suction cup volume, stiffness and elastic modulus have
a huge influence on the air leakage and hence the failure time of the cups. Based on our research we propose an
improved biomimetic design of suction cups, that could show improved failure times in varying degree of roughness
under dry and wet environments.
\end{abstract}

\maketitle

\setcounter{page}{1}
\pagenumbering{arabic}




{\bf 1 Introduction}

Contact mechanics of randomly rough solids involves many length scales, and quantities such as the real area of contact, 
the interfacial separation and the interfacial stress distribution are important for gaining insight in phenomenon 
such as adhesion, friction and leakage of seals\cite{Ref1,Ref2,Ref3,Ref4,Ref5,Ref6,Ref7,Ref8,Avi,BP,Creton,Gorb4,Mark}.

Flow of fluids, whether liquids or gases, through a rough interface from regions of high pressure to low pressure, 
is a technologically significant problem. Multiscale contact mechanics 
theories have been able to theoretically predict leak-rates in good agreement with experiment\cite{seal1,seal2,seal3,seal4,seal5}. 
In the simplest approach it is assumed that the full pressure drop occur 
over the most narrow constrictions (critical constrictions), along the biggest open (percolating) fluid flow channels.
If the contact area percolate no fluid leakage is expected.
When two elastic solids with randomly rough surfaces come in contact, numarical simulations show that the contact area 
percolate when the relative area of contact $A/A_0 \approx 0.42$\cite{Dapp}.

Leakage studies on static seals, dynamic seals and syringes have been a subject of 
many experimental and numerical studies in the past\cite{seal1,seal2,seal3,seal4,seal5,Dapp,BottiCarbone,Dapp2,martin1}. In this paper we study
the gas and water leakage in suction cups. Suction cups are used in ``simple'' applications, such as hanging items to smooth surfaces whether it be in cars or in houses,
and technologically complex applications such as robotic wall climbers which employ suction cups for their movement, which are used to
clean glass windows or to detect cracks in concrete structures, where human intervention is risky. 
In nature octopus vulgaris and the sucker fish have suction cups which they use for attaching themselves to surfaces varying in roughness. 
Currently, there is a lot of focus on developing suctions cups inspired by octopus which can attach itself to rough surfaces in dry and wet environments.

\begin{figure}
        \includegraphics[width=0.4\textwidth]{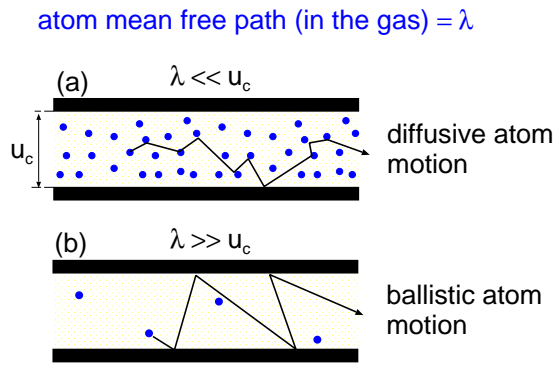}
        \caption{\label{Ballistic.eps}
Diffusive (a) and ballistic (b) motion of the gas atoms in the critical junction.
In case (a) the gas mean free path $\lambda$ is much smaller than the gap width $u_{\rm c}$ and the gas
molecules makes many collisions with other gas molecules before a collision with the solid walls.
In the opposite limit, when $\lambda >> u_{\rm c}$ the gas molecules makes many collisions with the solid wall before colliding
with another gas molecule. In the first case (a) the gas can be treated as a
(compressible) fluid, but this is not the case in (b). Adapted from \cite{arx}. }
\end{figure}

In this paper we study the leakage of air at the nominal contact region between a suction cup and the 
countersurface (substrate). We will show that for a suction cup to not fail within typical time of 
use ($\sim 1$ year) the surface separation $u_{\rm c}$ at the most narrow constrictions along the 
biggest percolating interfacial gas flow (leakage) channels must be below $2 \ {\rm nm}$. 
This is much smaller than the (average) air molecule mean-free-path $\lambda$ (due to collisions between the gas molecules), 
which implies that the gas flow between the collisions with the solid walls is mainly ballistic rather than diffusive (the so called Knudsen limit) (see Fig. \ref{Ballistic.eps}).

In Sec. 2, 3 and 4 we discuss concepts pertaining to diffusive and ballistic air leakage, and develope a simple theory for 
suction cups. In Sec. 5 we present experimental results 
and prediction results of air leakage into the suction cups. We show how the lifetime (time for deatachment) vary  with the
pull-off force and the surface roughness of counterface. In Sec. 5 we also discuss the effect of suction cup stiffness 
on its lifetime, along with discussion on influence of water leakage on suction cup lifetimes. In Sec. 6 we propose biomimetic 
design of a suction cup based on our research findings.

A short version of this paper has been submitted for consideration for publication elsewhere\cite{arx}.

\begin{figure}
        \includegraphics[width=0.45\textwidth]{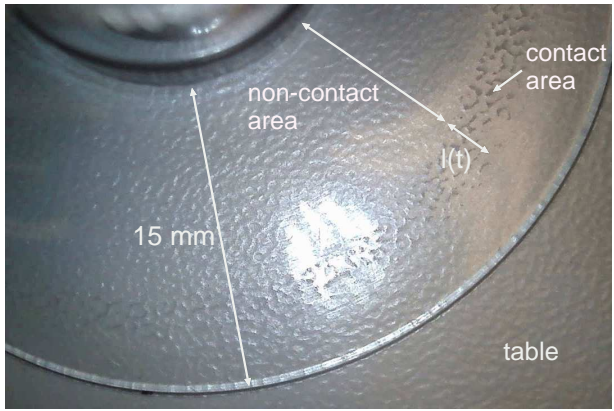}
        \caption{\label{pictureCONTACT.eps}
Suction cup squeezed against a corrugated (table) surface. Note that most of the area of real contact 
(darker regions) occur within an annular region of width $l \approx 2 \ {\rm mm}$,
formed close to the inner edge of the nominal contact area. We will denote this region as the contact region.
Note that the nominal contact pressure in the contact region appears to be nearly constant. 
In the present case the surface roughness on the substrate
is so high that the suction cup fails a few seconds after the normal squeezing force is removed.
The radial position of the contact region depends on the squeezing pressure, 
and moves towards the outer rim of the cup as the applied force decreases to zero.
}
\end{figure}

\vskip 0.3cm
{\bf 2 Qualitative discussion: diffusive or ballistic air flow?}

Consider a suction cup squeezed against a nominally flat surface and exposed to an external normal pulling force $F_1$.
A suction cup detach from the substrate after some time period $t$, which we will refer to as the failure time, 
due to leakage of air into the suction cup. For most suction cups and practical applications
the volume $V$ of gas at atmospheric pressure in the suction cup at the point of failure is of order $V \approx 1 \ {\rm cm}^3$. Here we assume
that gas leakage only occur in open (non-contact) channels at the rubber-substrate nominal contact area. For very smooth surfaces
the contact area may percolate resulting in no gas leakage, but here we assume some surface roughness and that percolating
non-contact channels exist. 

The rubber-substrate contact pressure is largest in a narrow annular region close to the inner contact radius
(see Fig. \ref{pictureCONTACT.eps}). Let $L_x$ be the
width of the contact region in the radial (gas-leakage) direction and $L_y = 2 \pi r$ the length of the contact annulus in the
angular direction. Typically $L_y/L_x \approx 20$. When the surface separation is small enough the gas-leakage is
given by the ballistic flow equation and the gas leak-rate (molecules per unit time) is (see Sec. 4.2):
$$\dot N_{\rm b} \approx {1\over 2} {L_y\over L_x} {p_{\rm a}-p_{\rm b} \over k_{\rm B} T} \bar v u_{\rm c}^2 . \eqno(1)$$ 
Here $p_{\rm a}$ is the external (atmospheric) pressure, and $p_{\rm b}$ the gas pressure inside the suction cup. 
The separation between the surfaces at the most narrow constrictions along the biggest open channels,
the so called critical junctions, is denoted by $u_{\rm c}$, and $k_{\rm B}T$ is the thermal energy.
The average velocity of an air molecule (at room temperature) is $\bar v \approx 500 \ {\rm m/s}$. We approximate (1) with
$$\dot N_{\rm b} \approx 10 {p_{\rm a}\over k_{\rm B} T} \bar v u_{\rm c}^2 \eqno(2)$$ 
The volume $V_{\rm b}$ of gas at atmospheric pressure contain $N_{\rm b}$ gas molecules where, 
assuming an ideal gas, $p_{\rm a} V_{\rm b} = N_{\rm b} k_{\rm B}T$. The time $t$ it takes for $N_{\rm b}$ gas molecules
to leak into the suction cup is (approximately) determined by $\dot N_{\rm b} t \approx N_{\rm b} = p_{\rm a} V_{\rm b} /k_{\rm B}T$. Using (2) this gives
$$V_{\rm b} \approx 10 \bar v u_{\rm c}^2 t$$
or
$$u_{\rm c} \approx \left ({V_{\rm b} \over 10 \bar v t} \right )^{1/2} . $$
Assuming $V_{\rm b} \approx 1 \ {\rm cm}^3$ and the failure time $t= 1 \ {\rm hour}$ we get $u_{\rm c} \approx 200 \ {\rm nm}$.
In a more realistic case, if a suction cup fail after one year ($t \approx 3\times 10^8 \ {\rm s }$) we get $u_{\rm c} \approx 2 \ {\rm nm}$.
Here we note that the gas leakage is mainly via the ballistic gas flow (as assumed above) when the gas molecule mean free path
$\lambda > 0.1 u_{\rm c}$ (see Sec. 4.3). Since the mean free path of a gas molecule at atmospheric pressure is of order $\approx 100 \ {\rm nm}$ it follows
that, even if a suction cup fail after just $\sim 1 \ {\rm hour}$, the gas molecules in the critical constriction move 
(mainly) ballistically, rather than diffusively, between the collisions with the solid walls. 
In some applications below the surface roughness is so large that 
the suction cup failure time (lifetime) is of order $100 \ {\rm s}$. 
In these cases the ballistic and diffusively gas transport are roughly equaly likely.

\begin{figure}
\includegraphics[width=0.35\textwidth,angle=0]{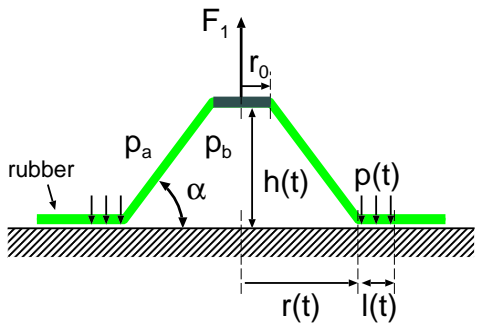}
\caption{\label{pic.eps}
Model of suction cup used in the present study. 
Adapted from \cite{arx}.
}
\end{figure}

\vskip 0.3cm
{\bf 3 Theory}

The suction cup we study below can be approximated as a trunkated cone with the diameter $2r_1$. The angle $\alpha$ 
and the upper plate radius $r_0$ are defined in Fig. \ref{pic.eps}.
When a suction cup is pressed in contact with a flat surface the rubber cone
will make apparent contact with the substrate in an annular region, but the contact pressure will be largest 
in a smaller annular region of width $l(t)$ formed close to
the inner edge of the nominal contact area (see Fig. \ref{pictureCONTACT.eps} and \ref{pic.eps}). 
We will assume that the rubber-substrate contact pressure in this region of space is constant, $p=p(t)$, and zero elsewhere. 

If we define $h_0 = r_0 {\rm tan} \alpha$ the volume of gas inside the suction cup is
$$V= \pi r^2 {1\over 3} (h+h_0) - \pi r_0^2 {1\over 3} h_0$$
Since 
$${r\over h_0 +h} = {r_0 \over h_0}$$
or
$${r\over r_0} = 1+{h \over h_0}\eqno(3)$$
we get
$$V = V_0 \left [\left ({r\over r_0}\right )^3-1\right ]\eqno(4)$$
where $V_0 = \pi r_0^2 h_0 /3$. 
The elastic deformation of the rubber film (cone) needed to make contact 
with the substrate require a normal force $F_0 (h)$, which we will refer
to as the cup (non-linear) spring-force. This force result from the bending of the film 
and to the (in-plane) stretching and compression of the film
needed to deform (part of) the conical surface into a flat circular disc or annulus
(see Appendix A).
We assume that the rubber cup is in repulsive contact with the substrate
over a region of width $l(t)$. Since the thickness of the suction cup material decreases as $r$ increases, we expect that $l$
decreases as $r$ increases. From optical pictures of the contact we have found that to a good approximation
$$l \approx l_0 +l_{\rm a} \left (1-{r\over r_0}\right ) = l_0 - l_a {h\over h_0}\eqno(5)$$
where $l_a = (l_1-l_0)/(1-r_1/r_0)$ where $l_1$ is the width of the contact region when $r=r_1$, 
and $l_0$ the width of the contact region when $r=r_0$.
The contact pressure $p=p(t)$ in the circular contact strip is assumed to be constant
$$p \approx { F_0(h) \over 2 \pi r l}+\beta (p_{\rm a}-p_{\rm b}), \eqno(6)$$
where $\beta$ is a number between 0.5 and 1, which depends on how the gas pressure 
change from $p_{\rm b}$ for $r=r(t)$ to $p_{\rm a}$ for $r=r(t)+l(t)$.

The function $F_0(h)$ can be easily measured experimentally (see below), or calculated 
using the FEM method (see also Appendix A). We note that
$F_0(h)$ depends on how fast the suction cup is compressed i.e. on the speed $\dot h(t)$. 
This is due to the viscoelastic nature of the suction cup material as will be discussed further below.

Assume that the pull-force $F_1$ act on the suction cup (see Fig. \ref{pic.eps}). The sum of $F_1$ and the
cup spring-force $F_0$ must equal the force resulting from the pressure difference
between outside and inside the suction cup,  i.e.
$$F_0+F_1 = \pi r^2  \left (p_{\rm a} - p_{\rm b}\right )\eqno(7)$$
We assume that the air can be treated as an ideal gas so that
$$p_{\rm b} V_{\rm b} = N_{\rm b} k_{\rm B}T.\eqno(8)$$

The number of molecules per unit time entering the suction cup, $\dot N_{\rm b} (t)$, is given by
$$\dot N_{\rm b} = f(p,p_{\rm a},p_{\rm b}) {L_y\over L_x}\eqno(9)$$
where 
$${L_y\over L_x} = {2 \pi r \over l}$$
The (square-unit) leak-rate function $f(p,p_{\rm a},p_{\rm b})$ will be discussed in Sec. 4.

The equations (3)-(9) constitute 7 equations from which the following 7 quantities can be obtained:
$h(t)$, $r(t)$, $l(t)$, $V(t)$, $p_{\rm b}(t)$, $p(t)$ and $N_{\rm b}(t)$.
The equations (3)-(9) can be easily solved by numerical integration. 

The suction cup stiffness force $F_0(h)$ depends on the speed with which the suction cup is compressed (or decompressed).
The reason for this is the viscoelastic nature of the suction cup material. To take this effect into account 
we define the contact time state variable $\phi (t)$ as\cite{Ruina,Rice,Ref1}:
$$\dot \phi = 1-\dot r \phi /l\eqno(10)$$
with $\phi(0) = 0$. 
For stationary contact, $\dot r = 0$, this equation gives just the time $t$ of stationary contact, $\phi (t) =t$.
When the ratio $\dot r/l$ is non-zero but constant (10) gives  
$$\phi (t) =  \left (1-e^{-t/\tau}\right ) \tau, $$
where $\tau = l/\dot r$. Thus for $t >> \tau$ we get $\phi (t) = \phi_0 = \tau$, which is the time a particular point on the suction cup surface
stay in the rubber-substrate contact region of width $l(t)$. It is only in this part of the rubber-substrate nominal contact region where a strong (repulsive)
interaction occur between the rubber film and the substrate, and it is region of space which is most important for the gas sealing process.

From dimensional arguments we expect that $F_0(h)$ is proportional to the effective elastic modulus of the cup material (see below and Appendix A).
We have measured $F_0(h)$ at a constant indentation speed $\dot h$, corresponding to a constant radial velocity $\dot r = \dot h (r_0/h_0)$
(see (3)). In this case the effective elastic modulus is determined by the relaxation modulus
$E_{\rm eff} (t)$ calculated for the contact time $\phi_0 = l/\dot r$. 
However, in general $\dot r$ may be strongly time-dependent. We can take that into account by 
replacing the measured $F_0(h)$ by the function $F_0(h) E_{\rm eff} (\phi(t))/E_{\rm eff} (\phi_0)$.

To calculate the leakrate function $f(p,p_{\rm a},p_{\rm b})$ we need the relation between the interfacial separation $u$ and the
contact pressure $p$. For this we have used the Persson contact mechanics theory\cite{BP,Alm,uc1}.
The relation between $p$ and $u$ depends on the elastic modulus of the suction cup. However, as pointed out above,
the suction cup is made from a viscoelastic material so the effective modulus is time (or frequency) dependent. 
In the present study we have taken this into account by using the relaxation modulus
evaluated at the time given by the contact state variable $\phi (t)$. This is a good approximation as long as the no slip occur
between the rubber and the substrate. When slip occur, depending on the slip velocity $v$ and the relevant surface roughness wavenumber
$q$, the suction cup material may be exposed to much higher perturbing frequencies $\omega \approx q v$ than the frequency $1/\phi(t)$ determined by contact state variable
$\phi(t)$. We will discuss this topic further below.

\vskip 0.3cm
{\bf 4 Diffusive and ballistic gas leakage}

The gas leakage result from the open (non-contact) channels at the interface between the rubber
film and the substrate. Most of the leakage occur in the biggest open flow channels. 
The most narrow constriction in the biggest open channels are denoted as the
critical constrictions. Most of the gas pressure drop occur over the critical constrictions,
which therefore determine the leak-rate to a good approximation.
The surface separation in the critical constrictions is denoted by $u_{\rm c}$. Theory shows that the lateral
size of the critical constrictions is much larger than the surface separation $u_{\rm c}$ (typically by
a factor of $\sim 100$)\cite{seal1,seal2,seal3}. 

In the theory for suction cups enters the leakrate function $f(p,p_{\rm a},p_{\rm b})$ (see (9)).
In this section we show how this function can be calculated when the gas flow through the critical constrictions occur
in the diffusive and ballistic limits (see Fig. \ref{Ballistic.eps}). We also present an interpolation formula which is (approximately) valid
independent of the ratio between the gas mean free path and the surface separation at the critical constrictions.

\vskip 0.2cm
{\bf 4.1 Diffusive flow}

We treat the air as an ideal gas so that the average gas molecule velocity 
$$\bar v = \left ( {8 k_{\rm B} T\over \pi m}\right )^{1/2} ,\eqno(11)$$
where $m$ is the mass of a gas molecule.
The mean free path $\lambda$ is the average distance an atom (or molecule) move (on the average) 
before it makes its first collision with another gas atom. The gas viscosity
$$\eta = {1\over 3} mn\bar v \lambda . \eqno(12)$$
where $n$ is the gas molecule number density.
Note that $\lambda \sim 1/n$ so that $\eta$ and $n \lambda$ are independent of the gas number density, and 
$n_{\rm a} \lambda_{\rm a} = n_{\rm b} \lambda_{\rm b}$.

If the gas can be treated as a compressible Newtonian fluid we get\cite{hydro}
$$\dot N_{\rm b} = {1  \over 24} {L_y\over L_x} {(p_{\rm a}^2 -p_{\rm b}^2 )\over k_{\rm B}T} {u_{\rm c}^3  \over \eta }\eqno(13)$$
Using $p_{\rm a} = n_{\rm a} k_{\rm B} T$,  $p_{\rm b} = n_{\rm b} k_{\rm B} T$ and (11) and (12) this gives
$$\dot N_{\rm b} = {\pi \over 64} {L_y\over L_x} \left ({n_{\rm a}\over \lambda_{\rm a}} -{n_{\rm b}\over \lambda_{\rm b}} \right ) \bar v u_{\rm c}^3$$

\begin{figure}
        \includegraphics[width=0.2\textwidth]{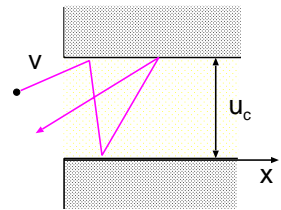}
        \caption{\label{BackReflacted.eps}
A gas atom which enter a constriction can be back scattered due to the atomic corrugation of the solid walls, or 
due to surface roughness or the thermal motion of the wall atoms.
}
\end{figure}

\begin{figure}
        \includegraphics[width=0.4\textwidth]{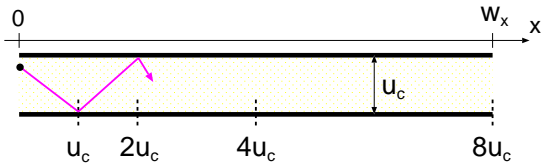}
        \caption{\label{Moving.eps}
An atom in the constriction collide with the walls in the channel at locations separated 
in the $x$-direction with on the average the distance $\approx u_{\rm c}$. We assume diffusive scattering of the
atom from the walls so when it is at $x=h$ (where it experience its first collision with a wall)
it has equal probability to return to $x=0$ as to move to
$x=2u_{\rm c}$. Hence the probability that it moves to $x=2u_{\rm c}$ is $1/2$. When it is at $x=2u_{\rm c}$ it has equal probability
to return to $x=0$ as to move to $x=4u_{\rm c}$. Hence the probability to move to $x=4u_{\rm c} = 2^2 u_{\rm c}$ will be $(1/2)\times (1/2) = (1/2)^2$.
In the same way, the probability to move to $x=2^n u_{\rm c}$ will be $(1/2)^n$. If it arrives to $x=w_x$ it is transferred
from the high pressure side to the low pressure side. Thus the probability that an atom which enter the
constriction at $x=0$ will exit it at $x=w_x$ will be $P=(1/2)^n$ where $2^n u_{\rm c}=w_x$. Hence $P\approx u_{\rm c}/w_x$. 
}
\end{figure}

\vskip 0.2cm
{\bf 4.2 Ballistic flow} 

We consider the flow of gas atoms through the critical junction in the ballistic limit $\lambda >> u_{\rm c}$.
This limit cannot be studied using continuum fluid mechanics (Navier Stokes equations), but require a
more atomistic approach, e.g., using the Boltzman equation\cite{Boltz1,Boltz2,Boltz3}. Here we will present a simple argument
(Fig. \ref{BackReflacted.eps} and \ref{Moving.eps}) which basically gives the result as a more accurate treatement.

We approximate the critical constriction with a rectangular junction with height $u_{\rm c}$ and width $w_y$ and length
(in the gas leakage direction) $w_x$.
The flow of atoms into the junction from the high pressure side is $(1/4) n_{\rm a} \bar v$ times the area
of the constriction given by $w_y u_{\rm c}$. 
(The factor of 1/4 is the result of the fact that only half of the atoms moves in the positive $x$-direction, and due to the fact
that on the average they only move in the $x$-direction with the speed $\bar v/2$.)
Thus the atom flow into the constriction is $(1/4) n_{\rm a} \bar vw_y u_{\rm c}$. However,
the gas atoms will scatter from the walls in a random way (diffusive scattering) due to atomic surface roughness on the walls, and also due to the
random thermal motion of the wall atoms. Thus some of the atoms which enter the junction will get back-scattered and return to the high 
pressure volume $V_{\rm a}$ (see Fig. \ref{BackReflacted.eps}). 
It turns out only a fraction $2 u_{\rm c}/w_x$ of the atoms will be able to penetrate the junction (see Fig. \ref{Moving.eps}). 
Thus the net number of atoms per unit time, leaving the constriction on the low pressure side 
after entering the constriction on the high pressure side, will be
$$\dot N_{\rm b} = {1\over 2} {w_y\over w_x} n_{\rm a} \bar v u_{\rm c}^2\eqno(14)$$ 
Taking into account the counter-flux of atoms (atoms, entering the critical constriction from the low pressure side, from the volume $V_{\rm b}$), this gives
$$\dot N_{\rm b} = {1\over 2} {w_y\over w_x} \left ( n_{\rm a} - n_{\rm b} \right ) \bar v u_{\rm c}^2\eqno(15)$$ 
In general, unless we are close to the point where the contact area percolate\cite{Dapp2,martin1}, $w_x \approx w_y$, and taking into account
the geometrical factor $L_y / L_x$ we get from (15)
$$\dot N_{\rm b} = {1\over 2}  {L_y\over L_x}  \left ( n_{\rm a} - n_{\rm b} \right ) \bar v u_{\rm c}^2\eqno(16)$$ 
Using the ideal gas law we can also write
$$\dot N_{\rm b} \approx {1\over 2} {L_y\over L_x} {p_{\rm a}-p_{\rm b} \over k_{\rm B} T} \bar v u_{\rm c}^2 \eqno(17)$$ 

\vskip 0.2cm
{\bf 4.3 Interpolation formula} 

We can interpolate between the limits (13) and (17) using
$$\dot N_{\rm b} = {1  \over 24} {L_y\over L_x} {(p_{\rm a}^2 -p_{\rm b}^2 )\over k_{\rm B}T} 
{u_{\rm c}^3  \over \eta }\left (1+12 {\eta \bar v \over (p_{\rm a}+p_{\rm b}) u_{\rm c}} \right )\eqno(18)$$
or
$$\dot N_{\rm b} = {1  \over 24} {L_y\over L_x} {(p_{\rm a}^2 -p_{\rm b}^2 )\over k_{\rm B}T} 
{u_{\rm c}^3  \over \eta }\left (1+\xi {\langle \lambda \rangle \over u_{\rm c}}\right )\eqno(19)$$
where $\xi = 32/\pi$ and where we define the average mean free path
$$\langle \lambda \rangle = {1\over 2} {n_{\rm a} \lambda_{\rm a} +n_{\rm b} \lambda_{\rm b}\over n_{\rm a} +n_{\rm b}}\eqno(20)$$
The gas leakage equation (19) is in good agreement with more accurate treatements using the Boltzman equation, and with experiments\cite{Boltz1}.

Note that $n_{\rm a} \lambda_{\rm a} = n_{\rm b} \lambda_{\rm b}$ so that if $n_{\rm a} >> n_{\rm b}$ we get $\langle \lambda \rangle \approx \lambda_{\rm a}$.
If $p_{\rm b} = 0$ we get from (19):
$$\dot N_{\rm b} = {\pi \over 64} {L_y\over L_x} u_{\rm c}^3 n_{\rm a} \bar v \left ( {1\over \lambda_{\rm a}} + {\xi \over u_{\rm c}}\right )\eqno(21)$$ 
We can interpret the factor
$${1\over \lambda_{\rm eff}} = {1\over \lambda_{\rm a}} + {\xi \over u_{\rm c}}\eqno(22)$$
as defining an effective mean free path. 
Using (19) we get
$$f(p,p_{\rm a},p_{\rm b}) = {1  \over 24} {(p_{\rm a}^2 -p_{\rm b}^2 )\over k_{\rm B}T} 
{u_{\rm c}^3  \over \eta }\left (1+\xi {\langle \lambda \rangle \over u_{\rm c}}\right )\eqno(23)$$
Note that $f$ depends on the contact pressure $p$ since the critical separation $u_{\rm c}$ depend on the contact
pressure.

\begin{figure}
        \includegraphics[width=0.35\textwidth]{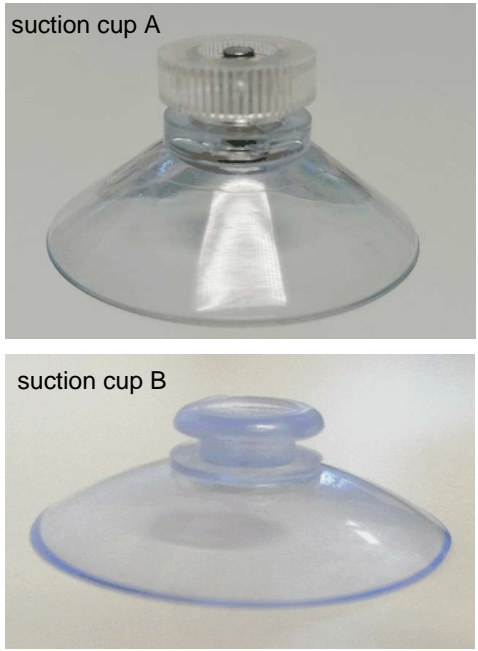}
        \caption{\label{SuctionCupPic1.eps}
The two suction cups A and B used in the present study. Both cups have the diameter 4 cm and are made from soft PVC.}
\end{figure}

\vskip 0.3cm
{\bf 5 Experimental results and analysis}

We will present experimental results for the suction cup A shown in Fig. \ref{SuctionCupPic1.eps}.
The suction cup is made from soft polyvinyl chloride (soft-PVC) and has a diameter 
of $2r_1 \approx 4 \ {\rm cm}$. We first present results for the
topography of the substrate surfaces used in the study. Next we present results for the relaxation
modulus of the soft-PVC, and the suction cup stiffness force. Finally we present measurements of the 
suction cup failure time under different conditions.


\begin{figure}
        \includegraphics[width=0.45\textwidth]{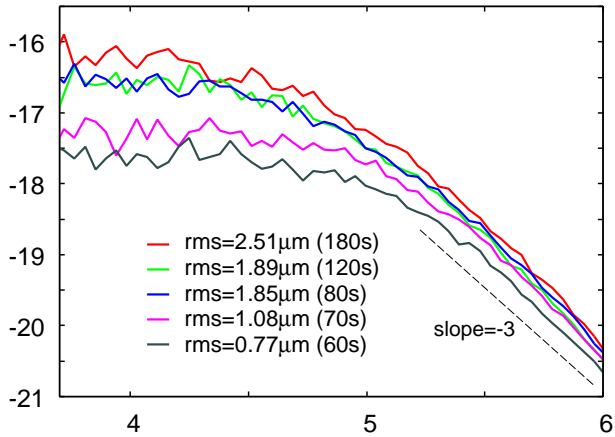}
        \caption{\label{1logq.2logC1D.19.17.23.22.20.eps}
Surface roughness power spectra of 5 sandblasted PMMA surfaces. The surfaces was sandblasted ``by hand'' under nominally identical conditions but
for different time periods, resulting in surfaces with different rms-roughness values. The rms-roughness and the sandblasting time
are indicated in the figure.}
\end{figure}

\begin{figure}
        \includegraphics[width=0.45\textwidth]{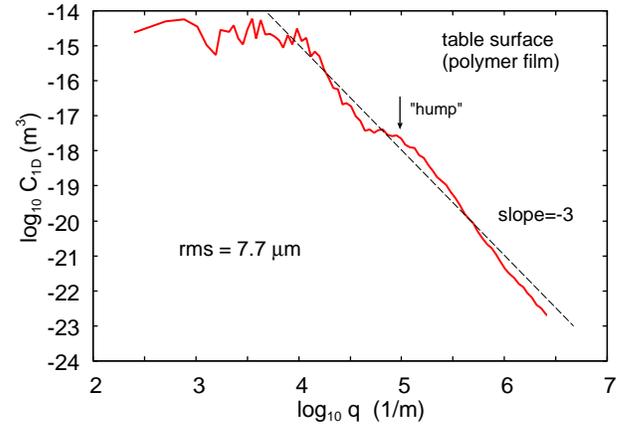}
        \caption{\label{1logq.2logC1D.BoTable.eps}
Surface roughness power spectra of table surface.}
\end{figure}

\begin{figure}
        \includegraphics[width=0.45\textwidth]{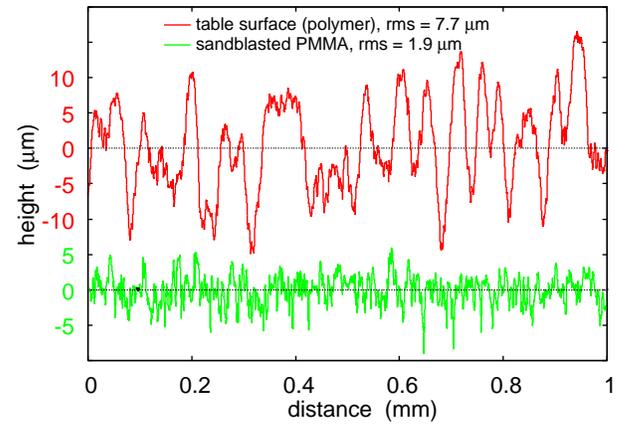}
        \caption{\label{1x.2h.Table.and.sandblasted120s.eps}
The surface height profiles of the table surface (red) and of sandblasted
PMMA (green). The surfaces have the rms roughness amplitudes $7.7 \ {\rm \mu m}$ and $1.9 \ {\rm \mu m}$, respectively.
Note that the table surface have a quasi-periodic structure with a wavelength of order $0.1 \ {\rm mm}$}
\end{figure}

\begin{figure}
        \includegraphics[width=0.45\textwidth]{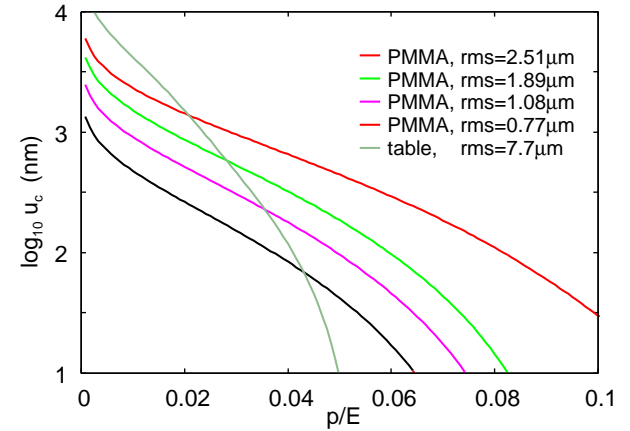}
        \caption{\label{1pressureMPa.2loguc.1.eps}
The logarithm of the separation $u_{\rm c}$ (in nm) at the critical constriction as a function of the applied
nominal squeezing pressure $p$ divided by the Young's modulus. }
\end{figure}

\vskip 0.2cm
{\bf 5.1 Surface roughness power spectra $C_{\rm 1D}$} 

The most important information about the surface topography of a rough surface is the surface roughness power spectrum.
For a one-dimensional (1D) line scan $z=h(x)$ the power spectrum is given by
$$C_{\rm 1D} (q) = {1\over 2 \pi} \int_{-\infty}^\infty dx \ \langle h(x) h(0) \rangle e^{i q x}$$
where $\langle .. \rangle$ stands for ensemble averaging.
For surfaces with isotropic roughness the 2D power spectrum $C(q)$ can be obtained directly from $C_{\rm 1D} (q)$ as described elsewhere \cite{Nyak,CarbLor}.
For randomly rough surfaces, all the (ensemble averaged) information about the surface is contained in the power spectrum $C(q)$. For this reason
the only information about the surface roughness which enter in contact mechanics
theories (with or without adhesion) is the function $C(q)$.
Thus, the  (ensemble averaged) area of real contact, the interfacial stress distribution and the
distribution of interfacial separations, are all determined by $C(q)$\cite{BP,Alm}.

Note that moments of the power spectrum determines standard
quantities which are output of most stylus instruments and often quoted.
Thus, for example, the mean-square (ms) roughness amplitude $\langle h^2 \rangle$
is given by
$$\langle h^2 \rangle = 2 \int_0^\infty dq \ C_{\rm 1D}(q).$$

Using an engineering stylus instrument we have studied the surface topography of 6 different surfaces used in our suction cup experiments.
The topography measurements was performed using Mitutoyo Portable Surface Roughness Measurement Surftest SJ-410 with a
diamond tip with the radius of curvature $R=1 \ {\rm \mu m}$, and with the tip-substrate repulsive force $F_{\rm N} = 0.75 \ {\rm mN}$.
The scan length $L=25 \ {\rm mm}$ and the tip speed $v=50 \ {\rm \mu m/s}$.

Fig. \ref{1logq.2logC1D.19.17.23.22.20.eps} shows the 1D surface roughness power spectra of 5 sandblasted poly(methyl methacrylate) (PMMA) surfaces. 
The surfaces was sandblasted under identical conditions but for different time periods resulting in surfaces with different rms roughness values
which increases from $0.77 \ {\rm \mu m}$ to $2.51 \ {\rm \mu m}$ as the sandblasting time increases from $60 \ {\rm s}$ to $180 \ {\rm s}$.

We also used a table surface (the surface shown in Fig. \ref{pictureCONTACT.eps}). The 1D power spectrum of this surface
is shown in Fig. \ref{1logq.2logC1D.BoTable.eps}. Note the ``hump'' in the power spectrum centered at $q \approx 10^5 \ {\rm m}^{-1}$ (indicated by an arrow in the figure),
corresponding to a wavelength of order $\lambda = 2 \pi /q \approx 0.1 \ {\rm mm}$. The surface structure generating this hump can be easily 
seen in Fig. \ref{pictureCONTACT.eps} and also detected in the measured line scan
height profile $h(x)$. Thus, in Fig \ref{1x.2h.Table.and.sandblasted120s.eps}
we show the surface height profiles of the table surface (red) and of a sandblasted
PMMA surface (green). The surfaces have the rms roughness amplitudes $7.7 \ {\rm \mu m}$ and $1.9 \ {\rm \mu m}$, respectively.
Note that the table surface have a quasi-periodic structure with a wavelength of order $0.1 \ {\rm mm}$,
while the sandblasted surface appears randomly rough.

Using the power spectra shown above we can calculate the height $u_{\rm c}$ of the critical constriction\cite{Alm,uc1,uc2}. 
This can be done directly using the critical junction theory\cite{seal3}, or by calculating first the leakrate using the effective medium theory\cite{seal2}
and identifying the calculated leakrate with the expected (critical junction) expression involving $u_{\rm c}$.
Both approaches gives nearly the same result but the second approach was used here as it is expected to be more accurate.

Fig. \ref{1pressureMPa.2loguc.1.eps}
shows the logarithm of the separation $u_{\rm c}$ (in nm) at the critical constriction as a function of the applied
nominal squeezing pressure $p$ divided by the Young's modulus. The most interesting pressure region is for $p$ of order or less than
1 atmosphere, which correspond to $p/E$ of order 0.025 or less (assuming $E \approx 4 \ {\rm MPa}$).
For the pressure 1 atmosphere the surface separation at the critical constrictions are 1.17, 0.68, 0.40 and $0.21 \ {\rm \mu m}$ for the sandblasted
PMMA with the rms roughness 2.51, 1.89, 1.08 and $0.77 \ {\rm \mu m}$. For the table surface at the same pressure $u_{\rm c} \approx 0.85 \ {\rm \mu m}$.
We note that for $u_{\rm c} \approx 0.5 \ {\rm \mu m}$ there is about equal probability for ballistic and diffusive gas flow, so clearly in the
present cases ballistic flow will contribute in an important way to the gas leakage into the suction cups.


\begin{figure}
        \includegraphics[width=0.45\textwidth]{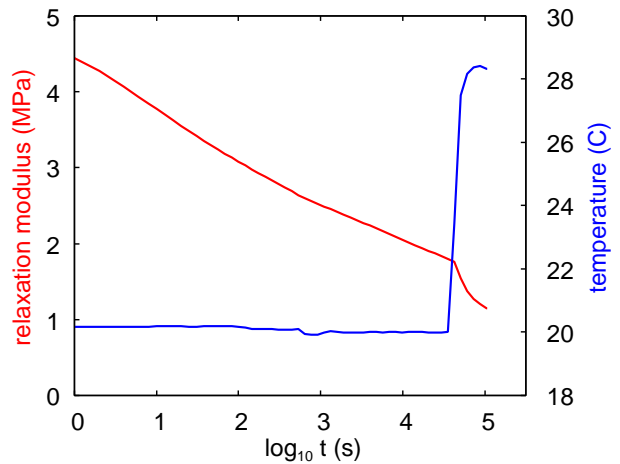}
        \caption{\label{relaxationSuctioncupmaterial01.eps}
The strain relaxation modulus $E_{\rm eff} (t) = \sigma_0/\epsilon(t)$ obtained with the applied
stress $\sigma_0 = 0.5 \ {\rm MPa}$. The temperature was kept fixed at $T=20^\circ {\rm C}$ for times
$t < 3\times 10^4 \ {\rm s}$ but was increased (due to run-out of the liquid nitrogen cooling fluid) 
to $T\approx 28^\circ {\rm C}$ at $t \approx 3 \times 10^4 \ {\rm s}$.}
\end{figure}

\vskip 0.2cm
{\bf 5.2 Viscoelastic relaxation modulus $E_{\rm eff}(t)$ } 
 
The two suction cups we use (see Fig. \ref{SuctionCupPic1.eps}) are made from a soft-PVC polymer.
Using a Q800 DMA instrument produced by TA Instruments, we have studied strain relaxation of the material used for cup A.
In Fig. \ref{relaxationSuctioncupmaterial01.eps} we show the effective Young's relaxation modulus $E_{\rm eff}(t)$
as a function of time $t$. In the experiments a thin strip (length $L_0$, cross section area $A_0$) of the polymer is loaded with a constant force $F$
in elongation mode at time $t=0$, and the length of the strip is measured as a function of time. 
From the strain $\epsilon(t)=[L(t)-L_0]/L_0$ we obtain the effective modulus $E_{\rm eff} = \sigma_0 /\epsilon(t)$ where $\sigma_0 = F/A_0$. 
We show results for $T=20^\circ {\rm C}$, where the strain change from $\approx 0.1$ to $\approx 0.4$ 
as the time increases from $1 \ {\rm s}$ to $t \approx 3 \times 10^4 \ {\rm s}$. Starting at  $t \approx 3 \times 10^4 \ {\rm s}$ the temperature gradually increased
to $T\approx 28^\circ {\rm C}$ with resulted in a faster increase in the strain. The effective Young's relaxation modulus decreases from $\approx 4.4 \ {\rm MPa}$
for $t=1 \ {\rm s}$ to $\approx 1.8 \ {\rm MPa}$ for $t \approx 3 \times 10^4 \ {\rm s}$.

We have also performed strain-sweep and temperature-sweep measurements for both suction cup A and B.
The results are presented in Appendix B and shows that the two suction cups are made from soft PVC with nearly the same
viscoelastic properties. Furthermore,  the stress-strain curve is nearly linear up to about $\epsilon \approx 0.3$,
which is in sharp contrast to rubber materials with filler which exhibit 
strongly non-linear properties already for the strain $\sim 0.001-0.01$ 
due to the break-up of the filler network\cite{lubricant}.

\begin{figure}
        \includegraphics[width=0.47\textwidth]{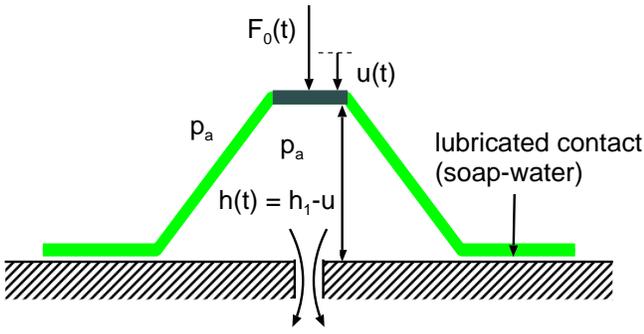}
        \caption{\label{pic.with.hole.eps}
Set-up used to measure the suction cup stiffness force curve. The force $F_0(t)$ to move the top of the
suction cup downward (or upward) with a constant velocity $\dot u$ is measured as a function of time
or displacement. A hole in the substrate surface allow the air to move out (or in during decompression),
and result in atmospheric pressure inside the suction cup. The rubber-substrate (here silica glass) is lubricated
by soap-water to remove the frictional shear stress at the interface.}
\end{figure}

\begin{figure}
        \includegraphics[width=0.45\textwidth]{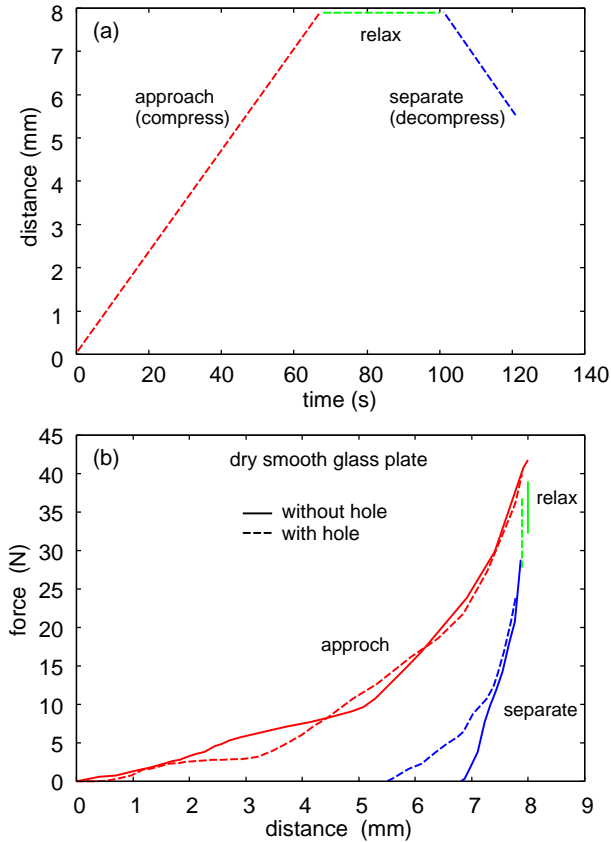}
        \caption{\label{1distance.2force.stiffness.Dry.with.without.hole.eps}
(a) Indentation cycle consisting of squeezing (approach, speed $v \approx 0.12 \ {\rm mm/s}$), relaxation ($v=0$ for $\Delta t \approx 35 \ {\rm s}$)
and retraction (separation or pull-off, speed $v \approx 0.12 \ {\rm mm/s}$).
(b) The interaction force (in N) acting on a suction cup as a function of the displacement (in mm).
The suction cup is squeezed against a smooth dry glass plate without (solid line) and with (dashed line) 
a hole in the center through which the air can leave so the air pressure
inside the rubber suction cup is the same as outside (atmospheric pressure).}
\end{figure}

\begin{figure}
        \includegraphics[width=0.45\textwidth]{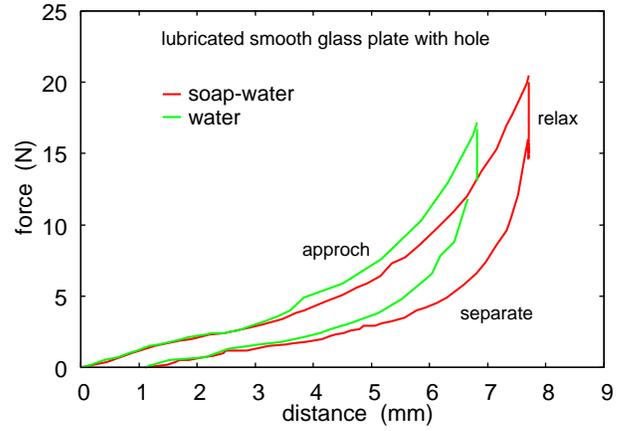}
        \caption{\label{1distance.2Force.Water.and.SoapWater.1.eps}
The force (in N) squeezing a suction cup as a function of the squeezing (or compression) distance (in mm).
The suction cup is squeezed against a smooth glass plate with a hole in the center through which the air can leave so the air pressure
inside the rubber suction cup is the same as outside (atmospheric pressure).The glass plate is lubricated with
water (green) and soap-water (red).}
\end{figure}

\begin{figure}
        \includegraphics[width=0.45\textwidth]{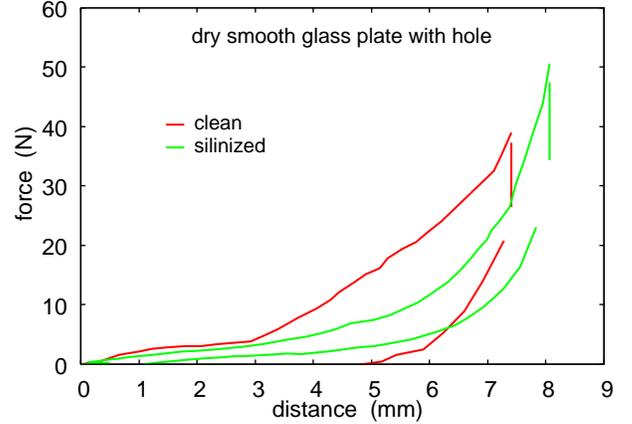}
        \caption{\label{1distance.2force.glass.hole.hydrophobic.eps}
The force (in N) squeezing a suction cup as a function of the squeezing (or compression) distance (in mm).
The suction cup is squeezed against a smooth glass plate with a hole in the center through which the air can leave so the air pressure
inside the rubber suction cup is the same as outside (atmospheric pressure). For clean dry glass (red) and dry silanized glass (green).}
\end{figure}

\begin{figure}
        \includegraphics[width=0.45\textwidth]{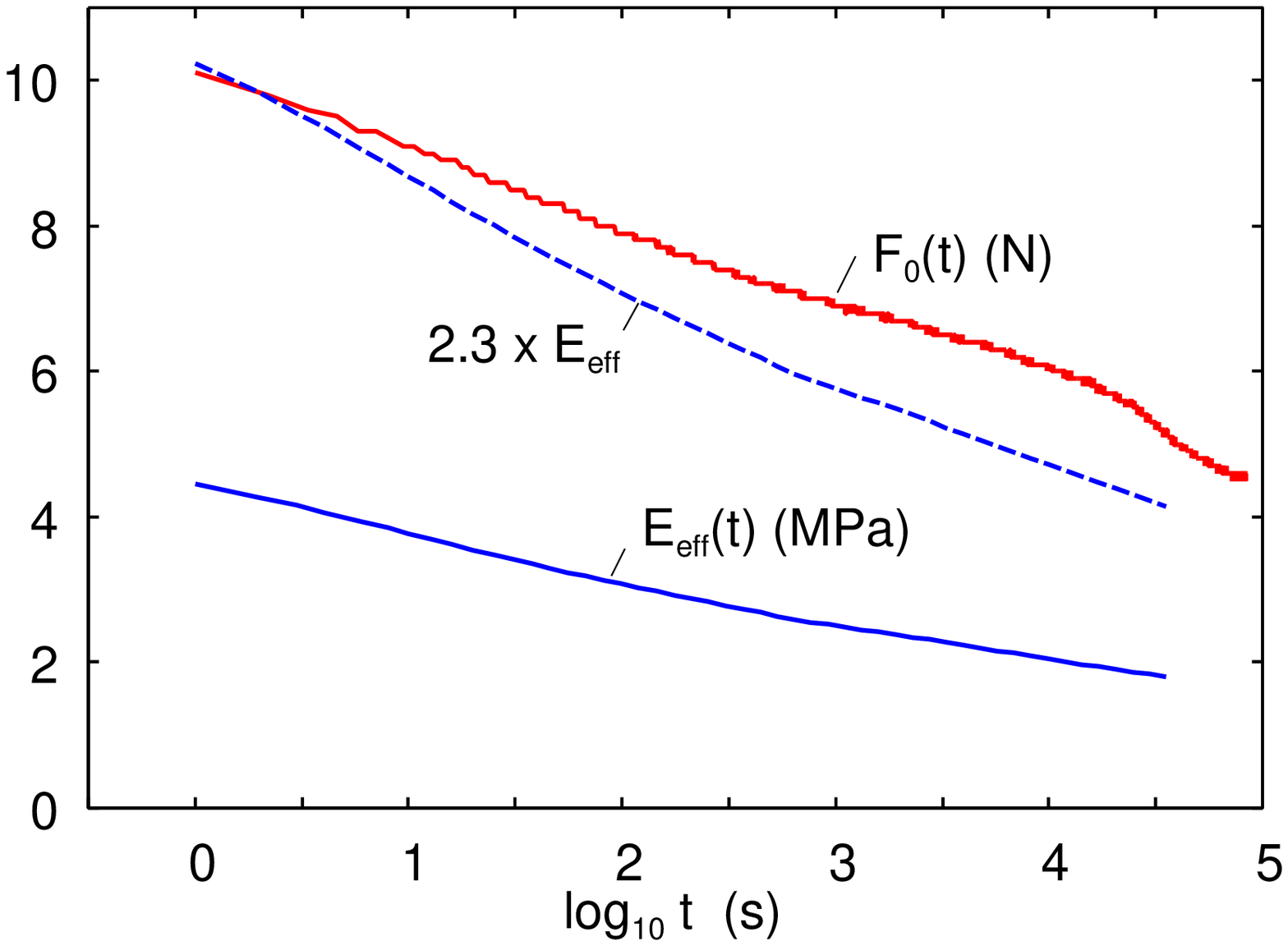}
        \caption{\label{1logt.2StiffnessForce.and.Erelax.eps}
Blue curve: The effective strain relaxation Young's modulus as a function of time as obtained in elongation mode at the applied stress $0.5 \ {\rm MPa}$.
Red curve: The suction cup stiffness force as a function of time. The dashed blue line is the relaxation modulus scaled by a factor of 2.3.}
\end{figure}

\vskip 0.2cm
{\bf 5.3 Suction cup stiffness force $F_0$} 

We have measured the relation between the normal force $F_0$ and the normal displacement of the (rigid) top plate of a suction cup.
In the experiments we increase the displacement of the top plate at a constant speed and measure the resulting force.
If $u$ denote the normal displacement,
where $u=0$ correspond to just before contact (undeform suction cup and $F_0=0$), 
then the height of the deformed suction cup $h=h_1-u$, 
where $h_1$ is the height of the undeformed suction cup (see Fig. \ref{pic.with.hole.eps}).

For the stiffness measurements we have used an instrument produced by SAUTER GmbH (Albstadt, Germany). 
The displacement can be changed with a resolution of 0.01 mm while the force sensor can measure forces up to 500 N with an accuracy of 0.1 N.

We have measured the relation $F_0(u)$ for 3 different cases were the suction cup was squeezed against (a) a glass plate lubricated by
soap water, (b) a dry clean glass plate and (c) a dry clean sandblasted glass plate. For case (a) and (b) we also performed measurements
when there was a hole in the in the glass plate (see Fig. \ref{pic.with.hole.eps}); this allowed the air to leave the suction cup
without any change in the pressure inside the suction cup (i.e. $p_{\rm b} = p_{\rm a}$ is equal to the atmospheric
pressure).

Fig. \ref{1distance.2force.stiffness.Dry.with.without.hole.eps} (b) shows 
the interaction force (in N) acting on a suction cup as a function of the displacement (in mm).
The suction cup is squeezed against a smooth dry glass plate without (solid line) and with (dashed line) 
a hole in the center through which the air can leave so the air pressure
inside the rubber suction cup is the same as outside (atmospheric pressure).
The indentation cycle [see Fig. \ref{1distance.2force.stiffness.Dry.with.without.hole.eps}(a)] 
consisting of squeezing (approach speed $v \approx 0.12 \ {\rm mm/s}$), 
relaxation ($v=0$ for $\Delta t \approx 35 \ {\rm s}$)
and retraction (separation or pull-off speed $v \approx 0.12 \ {\rm mm/s}$).
Note that during separation the suction cup pushed against the glass plate with the hole 
is able to retract farther than the suction cup pushed against the glass plate without a hole.
This is due to the fact that in the latter case (partial) vacuum exist inside the suction cup
and the top of the suction cup stop to retract when the spring force equal the vacuum force.

The result in Fig. \ref{1distance.2force.stiffness.Dry.with.without.hole.eps}(b) 
is influenced by the adhesion and friction between the suction cup and the glass plate. 
This is the reason for why even for the glass plate with a hole the
retraction stops for $u \approx 5.5 \ {\rm mm}$. However, the adhesion and friction can be removed (or strongly reduced)
if the glass surface is lubricated. This is illustrated in Fig. \ref{1distance.2Force.Water.and.SoapWater.1.eps}
which shows the interaction force as a function of the distance for the case of
a smooth glass plate (with a hole in the center) lubricated with water (green) and soap-water (red). In these cases
during retraction when $F_0=0$ the distance $u \approx 1 \ {\rm mm}$ i.e., the top of the suction cup has nearly returned to its
original position. The small displacement remaining (1 mm) may be attributed to viscoelastic relaxation of the suction cup
PVC material.

Adhesion and friction can also be reduced by coating the glass surface with an inert monolayer film.
This is illustrated in Fig. \ref{1distance.2force.glass.hole.hydrophobic.eps}
which shows the $F_0(u)$ curve for clean dry glass (red) and silanized glass (green).
Note that for the silanized glass the $F_0(u)$ curve is very similar to the case of the lubricated
glass surface.

\vskip 0.2cm
{\bf 5.4 Gas and water leakage}

We have studied how the failure time of a suction cup depends on the pull-off force (vertical load) and
the substrate surface roughness. We have also studied the influence 
on the failure time of a fluid (water) film at the rubber-substrate interface.
For the dry contact the suction cup was always attached to the lower side of a horizontal surface and a mass-load
was attached to the suction cup. We varied the mass-load from 0.25 kg to 8 kg.
If full vacuum would prevail inside the suction cup, the maximum possible pull-off force would be $\pi r_1^2 p_{\rm a}$.
Using $r_1 = 19 \ {\rm mm}$ and $p_{\rm a} = 100 \ {\rm kPa}$ we get $F_{\rm max} = 113 \ {\rm N}$ or about
11 kg mass load. However, the maximum load possible in our experiments for a smooth substrate surface 
is about 9 kg, indicating that no complete vacuum was obtained. This may, in least in part, 
be due to problems to fully remove the air inside the suction cup in the initial state. In addition we have
found that for mass loads above $8 \ {\rm kg}$ the pull-off is very sensitive 
to instabilities in the macroscopic deformations of the suction cup, probably
resulting from small deviations away from the vertical direction of the applied loading force. 

Below we will compare the experimental results to the theory predictions.
In the theory enter a few (mainly geometrical) parameters defined in Sec. 3,
which we have determined from simple measurements (see Table \ref{tab:RMS}).  

\begin{table}[hbt]
   \caption{Parameters used in the calculations. The (undeformed) suction cup upper and lower
radius are denoted by $r_0$ and $r_1$, respectively, and the suction cup angle by $\alpha$
(see Fig. \ref{pic.eps}). 
The initial radius (before gas-leakage into the cup) of the non-contact region 
(due to the spring force and due to air trapped inside the stopper,
but in the absence of an external loading force) is denoted by $r_{\rm start}$.
The width of the annular (high-pressure) cup-substrate contact region for $r=r_0$ and $r_1$
are denoted $l_0$ and $l_1$, respectively. The parameter $\beta$ determine the
contribution to the contact pressure from the pressure difference between inside and outside
the suction cup via $\Delta p = \beta (p_{\rm a}-p_{\rm b})$. Results are for suction cups A and B.
}
   \label{tab:RMS}
   \begin{center}
      \begin{tabular}{@{}|l||c|c|@{}}
         \hline
            quantity   &  cup A &  cup B \\
         \hline
         \hline
            $r_0$ & $5 \ {\rm mm}$  & $5 \ {\rm mm}$ \\
         \hline
            $r_1$ & $19 \ {\rm mm}$ & $19 \ {\rm mm}$ \\
         \hline
            $\alpha$ & $33^\circ$  & $21^\circ$ \\
         \hline
            $r_{\rm start}$ & $12 \ {\rm mm}$  & $12 \ {\rm mm}$ \\
         \hline
            $l_0$ & $4 \ {\rm mm}$  & $5 \ {\rm mm}$ \\
         \hline
            $l_1$ & $0.8 \ {\rm mm}$  & $2 \ {\rm mm}$ \\
         \hline
            $\beta $ & $0.8$  & $0.8$ \\
         \hline
      \end{tabular}
   \end{center}
\end{table}

\begin{figure}
        \includegraphics[width=0.45\textwidth]{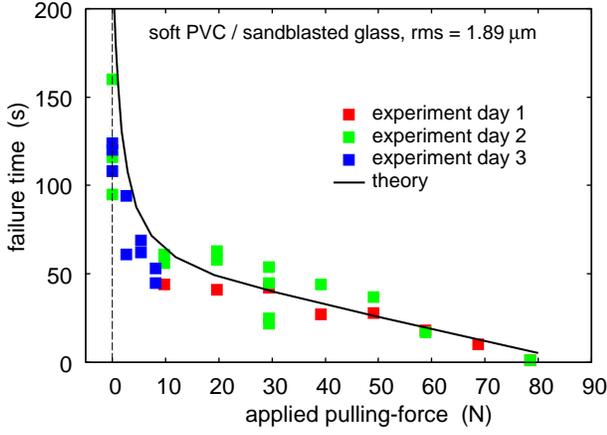}
        \caption{\label{1force.2time.xxx.eps}
The dependency of the pull-off time (failure time) on the applied (pulling) force. The soft PVC suction cup
is in contact with a sandblasted PMMA surface with the rms roughness $1.89 \ {\rm \mu m}$. Both surfaces were cleaned with
soap water before the experiments.
The symbols with different colors correspond to data obtained at different days.}
\end{figure}

\begin{figure}
        \includegraphics[width=0.45\textwidth]{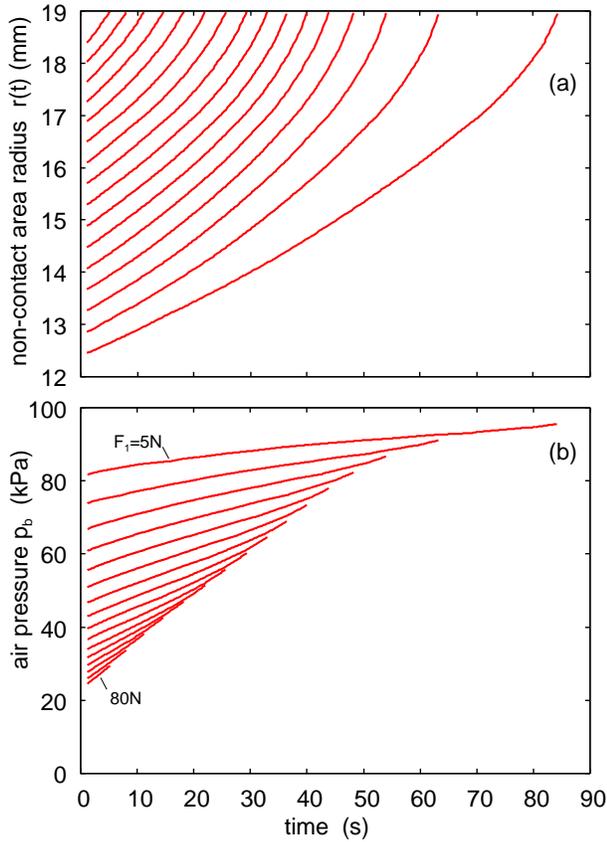}
        \caption{\label{1time.2r.3pa.x.eps}
The calculated dependency time dependency of the (a) radius of the non-contact region and (b) the pressure in the suction cup, 
for several pull-off forces (from $F_1=5 \ {\rm N}$ in steps of $5 \ {\rm N}$
to $80 \ {\rm N}$). The soft PVC suction cup is in contact with a sandblasted PMMA surface with the rms roughness $1.89 \ {\rm \mu m}$. 
Adapted from \cite{arx}.
}
\end{figure}

\begin{figure}
        \includegraphics[width=0.45\textwidth]{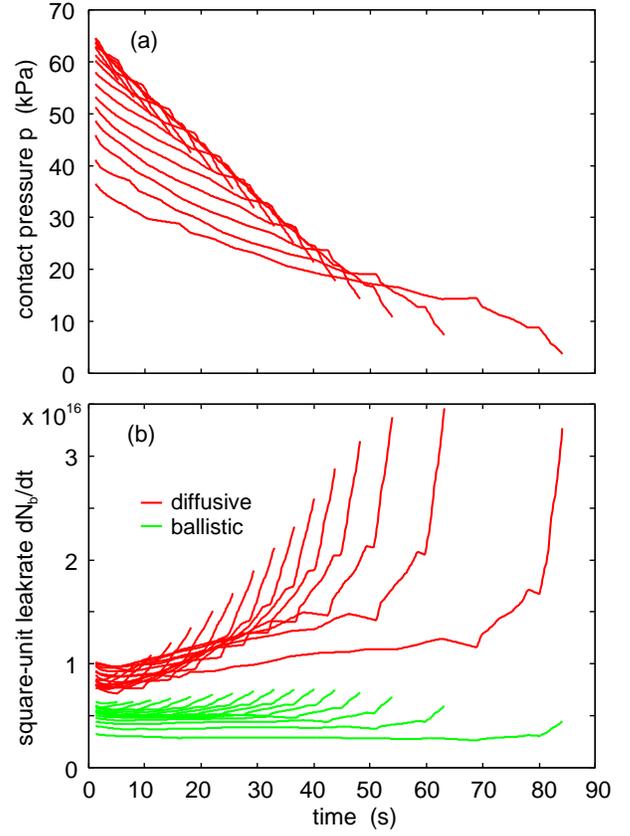}
        \caption{\label{1time.2p.dotN.x.eps}
The calculated the dependency of the (a) the contact pressure and (b) the square-unit gas leakage rate (molecules per unit time)
for several pull-off forces (from $F_1=5 \ {\rm N}$ in steps of $5 \ {\rm N}$
to $80 \ {\rm N}$). The soft PVC suction cup is in contact with a sandblasted PMMA surface with the rms roughness $1.89 \ {\rm \mu m}$. 
}
\end{figure}

\begin{figure}
        \includegraphics[width=0.45\textwidth]{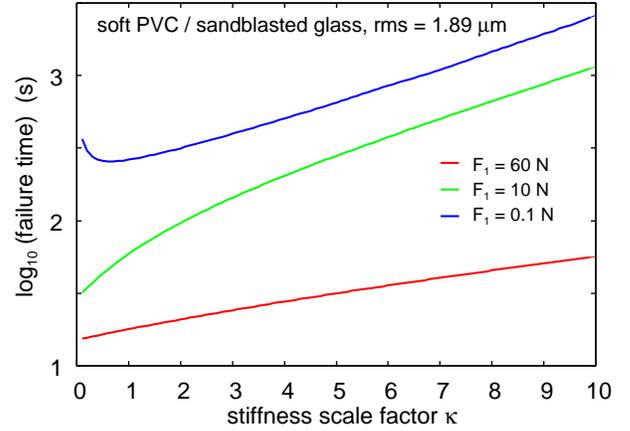}
        \caption{\label{1stiffnessFactor.2lifetime.eps}
The calculated stiffness dependency of the logarithm of the suction cup lifetime.
The results are for the same system as in Fig. \ref{1force.2time.xxx.eps} but we have scaled the stiffness force $F_0(h)$ by a factor $\kappa$.
That is, we have replaced $F_0(h)$ with $\kappa F_0(h)$, but all the other parameters are the same as in Fig. \ref{1force.2time.xxx.eps}. We show results for the
applied pull-off force $F_1=0.1 \ {\rm N}$ (blue), $10 \ {\rm N}$ (green) and $60 \ {\rm N}$ (blue).}
\end{figure}

\vskip 0.2cm
{\bf The dependency of the failure time on the pull-off force} 

Fig. \ref{1force.2time.xxx.eps} shows the dependency of the pull-off time 
(failure time) on the applied pulling force. The results are for the soft PVC suction cup
in contact with a sandblasted PMMA surface with the rms-roughness $1.89 \ {\rm \mu m}$. 
Before the measurement, both surfaces was cleaned with
soap water. The different colors of the square symbols correspond to data obtained at different days.
The solid line is the theory predictions, using as input the surface roughness power spectrum of the 
PMMA surface, and the measured stiffness of the suction cup, the latter corrected for viscoelastic time-relaxation
as described in Sec. 3. Note the good agreement between the theory and the experiments in spite of the simple nature
of the theory.

Fig. \ref{1time.2r.3pa.x.eps}(a) shows the calculated time dependency of the radius of the non-contact region, 
and (b) the pressure in the suction cup. We show results for several pull-off forces from $F_1=5 \ {\rm N}$ in steps of $5 \ {\rm N}$
to $80 \ {\rm N}$.  Fig. \ref{1time.2p.dotN.x.eps}(a) shows similar results for
the contact pressure and (b) for the square-unit gas leakage rate (molecules per unit time).
The noise in the curves reflect the noise in the measured stiffness and relaxation modulus curves which was used with linear interpolation, but without smoothing,
in the calculations. In the present case the air flow is mainly diffusively but for smoother surfaces the ballistic air flow dominate.

Finally, we study the influence of the suction cup stiffness on the lifetime.
Fig. \ref{1stiffnessFactor.2lifetime.eps} shows the calculated stiffness dependency of the logarithm of the suction cup lifetime.
The results are for the same system as in Fig. \ref{1force.2time.xxx.eps} but we have scaled the stiffness force $F_0(h)$ by a factor $\kappa$.
That is, we have replaced $F_0(h)$ with $\kappa F_0(h)$, but all other parameters are the same as in Fig. \ref{1force.2time.xxx.eps}. We show results for the
applied pull-off force $F_1=0.1 \ {\rm N}$ (blue), $10 \ {\rm N}$ (green) and $60 \ {\rm N}$ (blue). Note that increasing the magnitude of
$F_0(h)$ by a scaling factor result in a monotonic increase in the lifetime of the suction cup with increasing $\kappa$, except for very low applied pulling forces where
the lifetime first drop with increasing $\kappa$. The reason for the increase in the lifetime with increasing stiffness is the increase in the contact 
pressure $p$, and hence reduction in the surface separation $u_{\rm c}$ at the critical constrictions, with increasing stiffness.

\begin{figure}
        \includegraphics[width=0.45\textwidth]{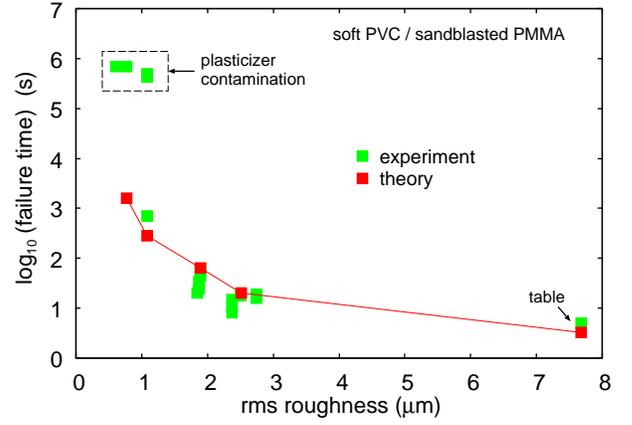}
        \caption{\label{1rms.2logFailureTime.eps}
The dependency of the pull-off time (failure time) on the substrate surface roughness. For soft PVC suction cups
in contact with sandblasted PMMA surfaces with different rms-roughness, and a table surface.
The pull-off force $F_1 = 10 \ {\rm N}$.
All surfaces were cleaned with soap water before the experiments, and a new suction cup was used for each experiment.
Adapted from \cite{arx}.}
\end{figure}

\begin{figure}
        \includegraphics[width=0.45\textwidth]{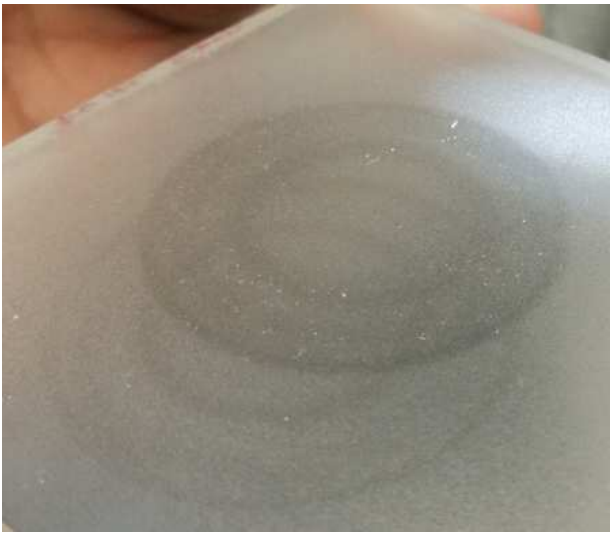}
        \caption{\label{Contamination.eps}
Footprint from a soft PVC suction cup after $\sim 10$ days contact. The suction cup and the PMMA surface was originally
cleaned with soap water. Mobile molecules (probably the PVC plasticizer)
have diffused from inside the PVC to the interface where it tend to fill-out the surface roughness
resulting in an effective smoother surface which scatter the light less than the uncovered surface. Note that there are two suction cup footprints. 
The weaker one resulted from an experiment performed several weeks earlier. This footprint is weaker than that of the last measurement 
because of evaporation of the contamination molecules and due to the cleaning action with soap 
water performed before the second measurement.}
\end{figure}

\vskip 0.2cm
{\bf The dependency of the failure time on the surface roughness} 

Fig. \ref{1rms.2logFailureTime.eps} shows the dependency of the pull-off time 
(failure time) on the substrate surface roughness. The results are for the type A soft PVC suction cups
in contact with sandblasted PMMA surfaces with different rms roughness, and a table surface.
All surfaces were cleaned with soap water before the experiments and a new suction cup was used for each experiment.
Note that for ``large'' roughness the predicted failure time is in good agreement with the measured data, but for rms roughness
below $\approx 1 \ {\rm \mu m}$ the measured failure times are much larger than the theory prediction. 
In addition, the dependency of the radius $r(t)$ of the non-contact region on time is very different in the two cases: 
For roughness larger than $\approx 1 \ {\rm \mu m}$ the radius increases 
continuously with time as also expected from theory (see Fig. \ref{1time.2r.3pa.x.eps}(a)).
For roughness below $\approx 1 \ {\rm \mu m}$  the boundary line $r(t)$ stopped to move a short time after applying the pull-off force, 
and remained fixed until the detachment occurred by a rapid increase in $r(t)$ (catastrophic event).
We attribute this discrepancy between theory and experiments 
to transfer of plasticizer from the soft PVC to the PVC-PMMA interface; this (high viscosity) 
fluid will fill-up the critical constrictions
and hence stop, or strongly reduce, the flow of air into the suction cup. This proposal is consistent with many studies of
soft PVC which shows relative fast migration of the plasticizer to the PVC surface. Thus, 
if $D$ is the diffusivity of a plasticizer molecule and $c$ the volume fraction of the plasticizer in the PVC,
then we expect that after time $t$ a layer of plasticizer of thickness $d \approx c (Dt)^{1/2}$ 
has diffused to the PVC-PMMA interface. The diffusivity $D$ depend on the type of plasticizer used, but at room temperature $D$ is
typically in the range\cite{plast1} $(1-100)\times 10^{-15} \ {\rm m^2/s}$. 
Assuming $c=1/2$ and $t=1000 \ {\rm s}$ we get $d\approx 1-10 \ {\rm \mu m}$. However, the concentrating of plasticizer may
be reduced in the surface region and in this case the transfer rate would be reduced. 
This is consistent with many studies\cite{plast2}
of the transfer of plasticizer from soft PVC to various contacting materials.
These studies show typical transfer rates
(at room temperature) correspond to a $\sim 1-10 \ {\rm \mu m}$ thick film of plasticizer 
after one week waiting time.  Additional support for the hypothesis that mobile molecules from the PVC block
the critical constrictions is the following observation:

Fig. \ref{Contamination.eps} shows the footprint from a soft PVC suction cup which detached after $\sim 10$ days of contact
with a sandblasted PMMA surface with the rms roughness $0.77 \ {\rm \mu m}$. 
In the experiment the suction cup was first squeezed into contact with the PMMA surface. Next 
a pull-off force (vertical load) was applied, resulting in the radius $r(t)$ increasing almost instantaneously
(in $\sim 1 \ {\rm s}$) to $r=r_*$, which is determined by the condition that the 
gas pressure force $(p_{\rm a} - p_{\rm b})\pi r_*^2$ equal the pull-off force plus the spring force. 
At this point the boundary line $r(t)$ stopped to move, and remained fixed (at $r=r_*$) until the detachment occurred by a rapid
(within $\sim 1 \ {\rm s}$) increase in $r(t)$ from $r_*$ to $r_1$.

The suction cup and the PMMA surface used in Fig.  \ref{Contamination.eps} was originally cleaned with soap water. 
However, Fig. \ref{Contamination.eps} shows that mobile molecules, probably the PVC plasticizer,
have diffused from inside the PVC to the interface where it tend to fill-out the surface roughness
cavities, resulting in an effective smoother surface which scatter the light less than the uncovered surface. 
We also observed that small dust particles tended to accumulate on the contaminated surface area, indicating that the 
contamination film is more sticky than the clean PMMA surface. 

The darkest (i.e. most contaminated) region in Fig. \ref{Contamination.eps} is the annular strip of width $l(t)$
where the contact pressure is highest during this waiting time period for $r(t)=r_*$. 
This is due to the high contact pressure in the region $r_* < r < r_*+l(t)$, which increases the transfer of the 
plastisizer to the PMMA surface as compared to the contact region $r_*+l(t) < r < r_1$, where the contact pressure is much smaller. 
This pressure-induced increase in the transfer of fluid (or mobile molecules) from a solid to the surface of 
a counter material is well known in the context of soft-PVC, and also know for other fluid filled materials, e.g., 
the cartilage of the human joints\cite{joint1,joint2,joint3}. 

Fig. \ref{Contamination.eps} shows two suction cup footprints. 
The weaker one resulted from an experiment performed several weeks earlier. This footprint is weaker than that of the last measurement 
because of evaporation of the contamination molecules, and due to the cleaning with soap 
water performed before the second measurement.

\begin{figure}
        \includegraphics[width=0.45\textwidth]{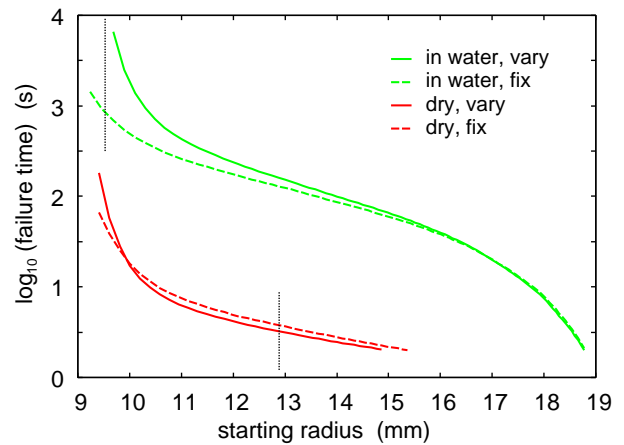}
        \caption{\label{1rStart.2logFailureTime.Table.eps}
The calculated dependency of the pull-off time (failure time) on the initial non-contact radius (for the pull-off force $F_1 = 10 \ {\rm N}$),
which depends on the force with which the suction cup was squeezed into contact with the counter surface (which determines the
amount of trapped gas or water in the suction cup in the initial state).
For soft PVC suction cups in contact with the table surface in dry state (red) and in water (green).
The solid lines use the relaxation modulus for calculating the interfacial separation while the dashed line use the effective modulus
after $1 \ {\rm s}$ contact time ($E=4 \ {\rm MPa}$). The vertical dotted lines indicate typical initial contact radius for the dry case and
in water.}
\end{figure}

\vskip 0.2cm
{\bf  Influence of a water film on the failure time} 

We have performed two experiments involving water. In the first experiment the substrate is covered by a thick water layer 
(about 1 cm thick), and in the second experiment we just added a few drops of water at the rim of the contact between the
suction cup and the substrate, after they had been squeezed together in the dry state. 
In these experiments the suction cup was squeezed against the
horizontal substrate from above, and a pulling force was applied to the suction cup. 

When a thick fluid film occur on the surface the suction cup failure time is much longer than in the
dry state. Assuming that the water can be treated as an incompressible Newtonian fluid, and assuming laminar flow,
the volume of fluid flowing per unit time into the suction cup is given by
$$\dot V_{\rm b} = {1  \over 12 \eta} {L_y\over L_x} u_{\rm c}^3   \left (p_{\rm a} -p_{\rm b} \right )\eqno(24)$$
The much higher viscosity of water than of air ($\approx 1\times 10^{-3} \ {\rm Pas}$ for water as
compared to $\approx 5\times 10^{-5} \ {\rm Pas}$ of air), result in a longer failure time in water than in air.
We have also found that in water the pull-off force is 
very sensitive to the applied squeezing force used in preparing the initial state. 
This is consistent with the theory. Thus, Fig. \ref{1rStart.2logFailureTime.Table.eps}
shows calculated results (for the table surface) for the dependency of the pull-off time (failure time) 
on the initial non-contact radius (for the pull-off force $F_1 = 10 \ {\rm N}$),
$r_{\rm start}$, in the dry state (red) and in water (green).
Note that $r_{\rm start}$ depends on the force with which the suction cup was squeezed into contact 
with the counter surface (which determines the amount of trapped gas or water in the suction cup in the initial state).

The vertical dotted lines in Fig. \ref{1rStart.2logFailureTime.Table.eps} 
indicate typical initial contact radius values for the dry case and
in water. Assuming the water can be treated as an incompressible fluid, the initial radius of the contact is smaller in water than in the dry state.
However, the low pressure in the trapped water can result in cavitation; this process depends on time 
and the amount of gas dissolved in the water, which makes the contact in water more complex than in the air.  
Visual inspection of suction cups squeezed against a smooth glass plate in water always show some trapped 
gas (and water) inside the suction cup. Nevertheless, the initial radius of the non-contact region is 
smaller when the contact is formed in water than if it is formed in the dry state.

The suction cup lifetime observed in water depends sensibly on relative 
small variations in the initial non-contact radius (resulting from fluctuations in
the force used in squeezing the suction cup in contact with the table surface). 
Thus, under nominally identical conditions in water the lifetime varied 
in the range $\approx 10^3-10^4 \ {\rm s}$. (Note: in all the experiments the suction cups were squeezed into contact 
with the substrate by a human hand, and some variations in the applied force cannot be avoided.) 
Similar experiments for dry conditions gives much smaller variations in the lifetime,
which vary in the range $\approx 3-6 \ {\rm s}$. This is consistent with Fig. \ref{1rStart.2logFailureTime.Table.eps}. 

In Fig. \ref{1rStart.2logFailureTime.Table.eps} the solid lines use the relaxation modulus $E_{\rm eff}(t)$ 
(evaluated for the time given by the state variable $\phi(t)$) when calculating the interfacial separation at the critical constriction. 
Using the relaxation modulus for calculating the time-dependency of the spring force (via the state variable $\phi(t)$) 
is a valid procedure, but if slip occur at the rubber-substrate interface then the same procedure may not be accurate for the
interfacial separation. Thus, the interfacial separation depends on the deformation frequencies the rubber is exposed to during slip 
between the suction cup and the counter surface. These frequencies depends on the slip velocity $v$ and on the 
wavelength $\lambda$ (or wavenumber $q=2\pi /\lambda$) of the relevant surface roughness components: $\omega \approx v q$. 
The most important roughness are the components with wavenumber smaller than the the wavenumber $q^*$ determined by the 
magnification $\zeta^*$ where the first non-contact percolating flow channel can be observed $q^* = \zeta^* q_0$ (see Ref. \cite{seal3}). 
Using the rubber friction theory developed elsewhere\cite{BP}, it is possible to take into account the dependency of the interfacial
separation on the sliding speed, but for the low slip velocities involved in most of the present applications (typically of order
$\sim 1 \ {\rm \mu m/s}$) the perturbing frequencies are low (of order $\sim 1 \ {\rm s}^{-1}$), and we expect only small modifications
of the results presented above. To illustrate this, the dashed lines in Fig. \ref{1rStart.2logFailureTime.Table.eps}  
use the effective modulus ($E_{\rm eff} \approx 4 \ {\rm MPa}$) corresponding to
$1 \ {\rm s}$ contact time (corresponding to the frequency $\omega \approx 1 \ {\rm s}^{-1}$). 

In the second experiment involving water the suction cup was first squeezed against the substrate in the dry state,
after which a few water drops was added at the contact rim. In this case we observed 
very violent flow processes (as in boiling of water), where non-periodic, pulsating flow of water-air mixtures
occurred in the open (non-contact) interfacial channels. In Ref. \cite{movie} we show a movie illustrating 
how mixtures of water and air is pulled into the suction cup via the (non-contact) leakage 
channels at the rubber-substrate interface.

\begin{figure}
        \includegraphics[width=0.45\textwidth]{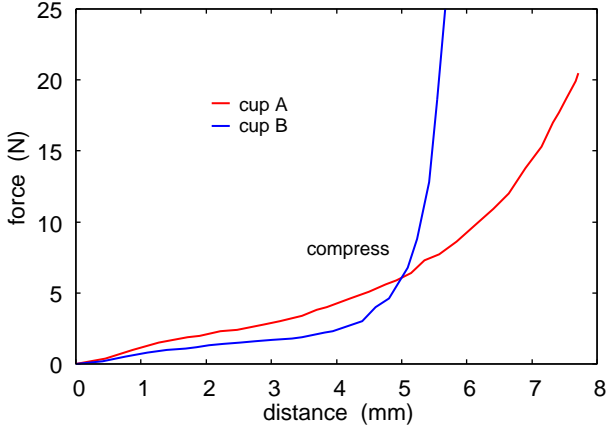}
        \caption{\label{1h.2F0.old.new.cup.eps}
The stiffness force $F_0(h)$ (in N) as a function of the squeezing (or compression) distance (in mm) for the suction cups A (red) and B (blue).
The suction cups are squeezed against a smooth glass plate with a hole in the center through which the air can leave so the air pressure
inside the rubber suction cup is the same as outside (atmospheric pressure).The glass plate is lubricated with soap-water.}
\end{figure}

\begin{figure}
        \includegraphics[width=0.45\textwidth]{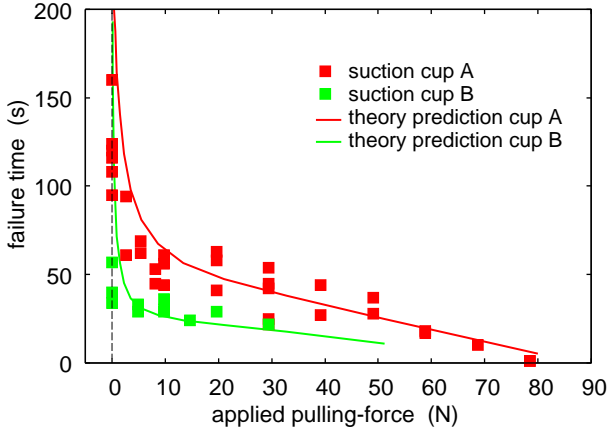}
        \caption{\label{1LoadingForce.2lifetime.oldRED.newGREEN.eps}
The dependency of the pull-off time (failure time) on the applied (pulling) force. The soft PVC suction cups A and B
are in contact with a sandblasted PMMA surface with the rms roughness $1.89 \ {\rm \mu m}$. All surfaces were cleaned with
soap water before the experiments.
Adapted from \cite{arx}.
}
\end{figure}

\vskip 0.2cm
{\bf 5.5 Results for another suction cup}

We have performed a few experiments for a second type of suction cup denoted by B (see Fig. \ref{SuctionCupPic1.eps}). The suction cups A and B 
are both made from soft PVC (see Appendix B) and both have the diameter $\approx 4 \ {\rm cm}$. 
However, for suction cup B the angle $\alpha = 21^\circ$ in contrast to $\alpha = 33^\circ$ for suction cup A.
In addition, the thickness $d$ of the PVC film of suction cup A is larger than for cup B. For both cups the thickness decreased nearly linarly 
with the distance $r$ from the center
of the suction cup (from $\approx 2.5 \ {\rm mm}$ at $r=5 \ {\rm mm}$ to $\approx 0.3 \ {\rm mm}$ at $r=20 \ {\rm mm}$ for cup A and 
from $\approx  1.5 \ {\rm mm}$ at $r=5 \ {\rm mm}$ to $\approx 0.3 \ {\rm mm}$ at $r=20 \ {\rm mm}$ for cup B). 
This difference in the thickness variation influence
the suction cup stiffness force as shown in Fig. \ref{1h.2F0.old.new.cup.eps}. Note that before the strong increase in the $F_0(h)$ curve
which result when the suction cup is squeezed into complete contact with the counter surface (and is due to compression of the suction cup material)
the suction cup A has a stiffness nearly twice as high as that of the suction cup B. 
Note that the fact that the angle $\alpha$ is smaller for the cup B than for the cup A 
is the reason for why a smaller displacement is needed in order to reach the point where
the stiffness force abruptly increases (where the suction cup is in complete contact with the substrate), see Fig. \ref{1h.2F0.old.new.cup.eps}.

The smaller angle $\alpha$ for suction cup B than for cup A imply that if the same amount of gas would leak into the suction cups the
gas pressure $p_{\rm b}$ inside the suction cup will be highest for the suction cup B. This will tend to reduce the lifetime of cup B. Similarly,
the smaller stiffness of the cup B result in smaller contact pressure $p$, 
which will increase the leakage rate and reduce the lifetime (see Fig. \ref{1stiffnessFactor.2lifetime.eps}).
Hence both effects will make the lifetime of the suction cup B smaller than that of the cup A.

In Fig. \ref{1LoadingForce.2lifetime.oldRED.newGREEN.eps} we show the dependency of the pull-off time (failure time) on the applied (pulling) force for 
suction cup B for the sandblasted PMMA surface with the rms roughness $1.89 \ {\rm \mu m}$ (green squares and line). 
We also show the results for suction cup A (red) (from Fig. \ref{1force.2time.xxx.eps}). In the calculations we have used the parameters shown in table \ref{tab:RMS}.
As expected, the lifetime of the suction cup B is less than that of suction cup A.

\begin{figure}
        \includegraphics[width=0.45\textwidth]{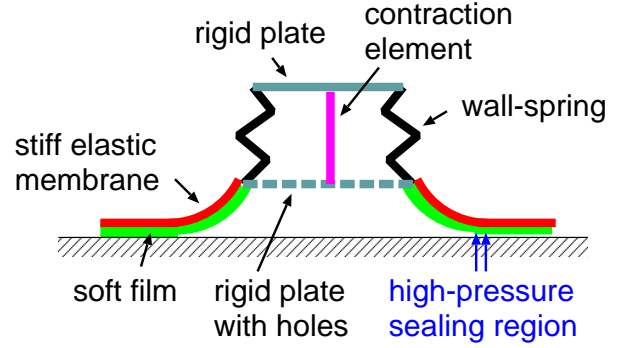}
        \caption{\label{SuctionCupMechanical.eps}
Octopus inspired suction cup. The suction cup consist of a soft material (green), e.g., double network hydrogel,
covered by elastically stiff membrane (red) in order to increase the contact stiffness and hence the contact 
pressure in the sealing region. Without the
contraction filament (pink) the suction cup must be squeezed in contact with the counter surface. With the contraction filament,
which could be a muscle fiber, the suction cup body (black) can be 
compressed before contact with the substrate. After contact with the substrate the
compression filament is relaxed resulting in an elongation of the suction cup body and a reduction in the water pressure inside the
suction cup. As a result, the suction cup will be compressed against the substrate.}
\end{figure}

\vskip 0.3cm
{\bf 6 Discussion}

Many animals have developed suction cups to adhere to different surfaces, e.g., the octopus\cite{Smith,Gorb1,Gorb2,Benne} or
northern clingfish\cite{Gorb3}. Studies have shown that these animals can adhere to much rougher
surfaces then man-made suction cups. This is due to the very low elastic modulus of the material
covering the suction cup surfaces. Thus, while most man-made suction cups are made from
rubber-like materials with a Young's modulus of order a few MPa, the suction cups of the octopus and the northern clingfish
are covered by very soft materials with an effective modulus of order $10 \ {\rm kPa}$. 

When a block of a very soft material is squeezed against a counter surface in water it tends to trap islands of water
which reduces the contact area and the friction\cite{Mug1,Mug2,Mug3}. For this reason 
the surfaces of the soft adhesive discs in the octopus and the northern clingfish 
have channels which allow the water to get removed faster during the approach 
of the suction cup to the counter surface. However, due to the low elastic modulus of the suction cup material,
the channels are ``flattened-out'' when the suction cup is in adhesive contact with the counter surface, and negligible
fluid leakage is expected to result from these surface structures.

There are two ways to attach a suction cup to a counter surface. Either a squeezing force is applied, or a
pump must be used to lower the fluid (gas or liquid) pressure inside the suction cup. The latter is used in some engineering applications.
However, it is not always easy for the octopus to apply a large normal force when attaching a suction cup to a counter surface,
in particular before any arm is attached, and when the animal cannot wind the arm around the counter surface as may be the 
case in some accounts with sperm whale.
Similarly, the northern clingfish cannot apply a large normal force to squeeze the adhesive disk in contact with a counter
surface. So how can they attach the suction cups? We believe it may be due to changes in the suction cup volume involving muscle contraction
as illustrated in the octopus inspired suction cup shown in Fig. \ref{SuctionCupMechanical.eps}.

The suction cup in Fig. \ref{SuctionCupMechanical.eps} consist of a soft material (green), e.g., double network hydrogel\cite{Gong},
covered by elastically stiff membrane (red) in order to increase the contact stiffness, and hence the contact 
pressure in the sealing region. Without the
contraction filament (pink) the suction cup must be squeezed in contact with the counter surface. With the contraction filament,
the suction cup body (black) can be compressed before contact with the substrate. After contact with the substrate the
compression filament is relaxed resulting in an elongation of the suction cup body and fluid flow into the cup body via the
holes in the bottom plate. This result in a reduction in the water pressure inside the
suction cup. As a result, the suction cup will be compressed against the substrate.

For adhesion to very rough surfaces, the part of the suction cup in contact with the substrate 
must be made from an elastically very soft material. 
However, using a very soft material everywhere result in a very small suction cup stiffness. We have shown 
(see Fig. \ref{1stiffnessFactor.2lifetime.eps}) that a long lifetime require a large enough suction cup stiffness. 
Only in this case will the contact pressure $p$ be large enough to reduce the water
leakrate to small enough values. For this reason we suggest that the soft material (green) is bound to a stiffer
material (red) as indicated in Fig. \ref{SuctionCupMechanical.eps}. The thickness of the green layer must be at
least of order the longest wavelength of the important surface roughness on the counter surface in order to be able to
deform at the interface and make nearly complete contact with the counter surface.
   
For suction cups used in the air the lowest possible pressure inside the suction cup is $p_{\rm b} = 0$, 
corresponding to perfect vacuum. In reality it is usually much larger, $p_{\rm b} \approx 30-90 \ {\rm kPa}$. For suction cups
in water the pressure inside the suction cup could be negative where the liquid is under mechanical tension.
This state is only metastable, but pressures as low as $p_{\rm b} \approx - 20 \ {\rm MPa}$ has been observed
for short times\cite{Cav}. For water in equilibrium with the normal atmosphere one expect cavitation to occur for any pressure below
the atmospheric pressure, but the nucleation of cavitives may take long time, and depends strongly on impurities
and imperfections. For example, crack-like surface defects in hydrophobic materials may trap small (micrometer or nanometer)
air bubbles which could expand to macroscopic size when the pressure is reduced below the atmospheric pressure.

Negative pressures have been observed inside the suction cups of octopus. Thus, in one study\cite{Smith1} it was found that
most suction cups have pressures above zero, but some suction cups showed pressures as low as $\approx -650 \ {\rm kPa}$.

Trapped air bubbles could influence the water flow into the suction cup by blocking flow channels.
For not degassed water, whenever the fluid pressure falls below the atmospheric pressure cavitation can occur, 
and gas bubbles could form in the flow channels and block the fluid flow due to the Laplace pressure effect. 
For the suction cups studied above, in the initial state the Laplace pressure is likely to be smaller than the fluid pressure difference
between inside and outside the suction cup, in which case the gas bubbles would get removed, but at a later stage in the
detachment process this may no longer be true. This is similar to observations in earlier model studies of the water leakage of
rubber seals, where strongly reduced leakage rates was observed for hydrophobic surfaces when the  
water pressure difference between inside and outside the seal become small enough\cite{seal4}.

\vskip 0.3cm
{\bf 7 Summary and conclusion}

We have studied the leakage of suction cups both experimentally and using a multiscale contact mechanics theory.
In order to describe the air leakage through the contact between a suction cup
and randomly rough counterface, a theory was developed which describes the air flow in both the 
diffusive limit ($\lambda<< u\textsubscript{c}$) and ballistic limit ($\lambda >> u\textsubscript{c}$),
where $u_{\rm c}$ is the interfacial separation at the critical constriction.
In the experiments suction cups (made of soft PVC) was pressed against sandblasted 
PMMA sheets. We found the failure times of suction cups to be in good agreement with the theory, 
except for surfaces with rms-roughness below $\approx 1 {\rm \mu m}$, where diffusion of plasticizer 
from the PVC to the PMMA counterface resulting in blocking of critical contrictions. 
The study shows that the the suction cup volume and suction cup stiffness are 
important design parameters, but most important is the elastic modulus of the suction cup material;
a low modulus can drastically help in reducing gas (or liquid) leakage.
We have suggested an improved biomimetic design of suction cups, inspired by octopus vulgaris, 
which could show improved failure time under water and on surfaces with varying degree of surface roughness.

\begin{figure}
        \includegraphics[width=0.45\textwidth]{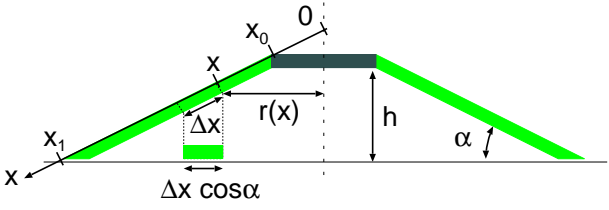}
        \caption{\label{ElasticEnergy.eps}
The elastic energy stored in the suction cup when it is pressed against the substrate is assumed to
be given by the work nexcessary to compress the suction cup in the radial direction so that each radial
segment of length $\Delta x$ is projected vertically upon the flat substrate surface and result in a segment of
width $\Delta x \ {\rm cos}\alpha$.}
\end{figure}

\vskip 0.3cm
{\bf Appendix A: Estimation of the stiffness force $F_0$}

We will estimate the stiffness force $F_0$ assuming there is no-slip at the rubber-substrate interface.
The no slip assumption holds approximately for the suction cup in contact with the smooth clean glass plate with a hole. 
The elastic energy stored in the suction cup when it is pressed against the substrate is assumed to
be given by the work nexcessary to compress the suction cup in the radial direction so that each radial
segment of length $\Delta x$ is projected vertically upon the flat substrate surface and result in a segment of
length $\Delta x \ {\rm cos}\alpha$ (see Fig. \ref{ElasticEnergy.eps}). The radial strain in the compressed
segment is
$$\epsilon = {\Delta x - \Delta x {\rm cos} \alpha \over \Delta x} = 1-{\rm cos} \alpha$$
The elastic energy
$$U= {1\over 2} V E \epsilon^2 = {1\over 2} V E (1-{\rm cos} \alpha)^2$$
where $E$ is Young's modulus and $V$ the volume of deformed material. If we have squeezed the region from $x$ to $x_1$
into contact with the substrate the volume
$$V= \int_x^{x_1} dx \ (2 \pi x {\rm cos} \alpha) d(x)$$
where $d(x)$ is the thickness of the suction cup at the radial location $x$.
If we assume $d(x)= a+bx$, where $a=(d_0 x_1 - d_1 x_0)/(x_1-x_0)$ and $b=(d_1-d_0)/(x_1-x_0)$ we get
$$V= 2 \pi \left [ {1\over 2} a (x_1^2-x^2) +{1\over 3} b (x_1^3-x^3)\right ]{\rm cos} \alpha $$
To calculate the normal force (in response to the normal displacement $u$) we use that
$${dV\over du}= {dV \over dx} {dx \over du}$$
Since
$$u=(x_1-x) {\rm sin} \alpha$$
we get 
$${dV\over du}= 2 \pi {{\rm cos} \alpha  \over {\rm sin} \alpha} \left (ax+bx^2 \right )$$
and
$$F_0 =  {dU\over du} = \pi (1-{\rm cos} \alpha)^2 {{\rm cos} \alpha \over {\rm sin} \alpha} E \left (ax+bx^2 \right )$$ 
As an example, assume $d_1=0$ so that $a=d_0 x_1/(x_1-x_0)$ and $b=-d_0/(x_1-x_0)$ we get
$$F_0 = \pi (E d_0) (1-{\rm cos} \alpha)^2 { {\rm cos} \alpha \over {\rm sin}\alpha} {x (x_1 -x) \over x_1-x_0}$$ 
Using that $x_1-x= u/{\rm sin} \alpha$ this gives
$$F_0 = \pi (E d_0) (1-{\rm cos} \alpha)^2 { {\rm cos} \alpha \over {\rm sin}^2\alpha}  {u (x_1-u/{\rm sin} \alpha )\over x_1-x_0}$$ 
For small indentation $u$ this gives a stiffness force which increase linearly with $u$. Using
$E= 4 \ {\rm MPa}$, $d_0 = 2 \ {\rm mm}$ and $\alpha = 33^\circ$ (cup A) and measuring $u$ in mm this eqation gives
$F_0 \approx (2.5 \ {\rm N/mm}) u$ which is in good agreement with the measured stiffness 
for dry contact (see Fig. \ref{1distance.2force.stiffness.Dry.with.without.hole.eps}) for $u$ up to a few about ${\rm mm}$
(note $u/{\rm sin} \alpha = x_1$ for $u \approx 11 \ {\rm mm}$). For $u$ above $\approx 5 \ {\rm mm}$ 
the compression of the rubber will start to contribute to $F_0$, and $F_0$ will increase more rapidly with $u$
then predicted by the theory above. 

\begin{figure}
        \includegraphics[width=0.45\textwidth]{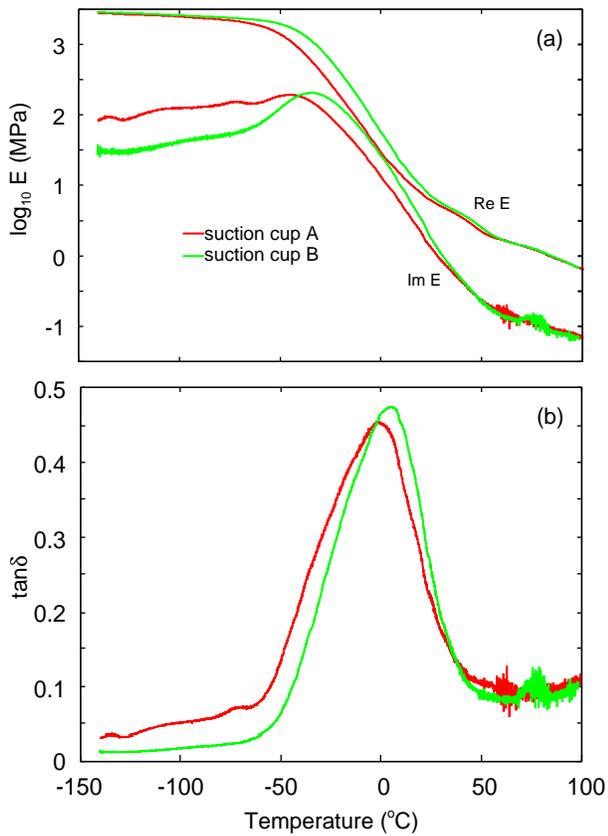}
        \caption{\label{ModlusVsTemperature.and.tand.eps}
The dependency of the low strain ($\epsilon = 0.0004$) viscoelastic modulus (a), and ${\rm tan}\delta = {\rm Im}E/{\rm Re}E$ (b), 
on the temperature for the frequency $f=1 \ {\rm Hz}$. For the soft PVC of suction cup A (red) and B (green).}
\end{figure}

\begin{figure}
        \includegraphics[width=0.45\textwidth]{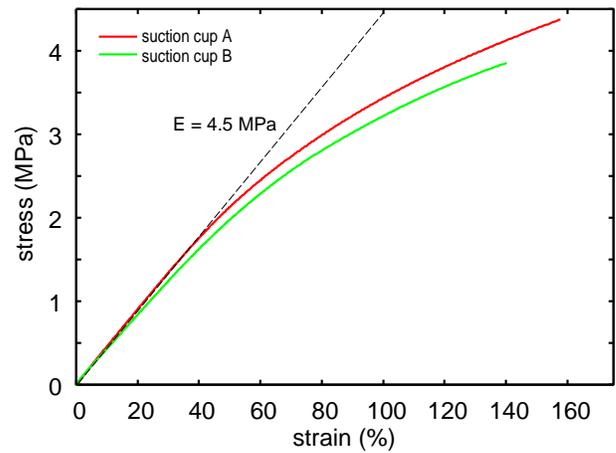}
        \caption{\label{stressStrain.eps}
The dependency of the stress on the strain for the soft PVC used for the suction cup A (red) and B (green).
The strain was increased from zero to its final value in about $300 \ {\rm s}$ giving a strain rate of
about $0.003 \ {\rm s^{-1}}$.}
\end{figure}

\vskip 0.3cm
{\bf Appendix B: Viscoelastic modulus of the soft PVC}

We have measured the viscoelastic modulus of the soft PVC of the suction cups A and B. The measurements where performed  
in oscillatory tension mode using a Q800 DMA instrument produced by TA Instruments.
Fig. \ref{ModlusVsTemperature.and.tand.eps} shows (a) the temperature dependency of the low strain 
($\epsilon = 0.0004$) modulus $E$, and (b) ${\rm tan}\delta = {\rm Im}E/{\rm Re}E$, 
for the frequency $f=1 \ {\rm Hz}$. Results are shown for the soft PVC of suction cup A (red) and B (green).
Note that both materials exhibit very similar viscoelastic modulus.
If we define the glass transition temperature as the temperature where
${\rm tan}\delta$ is maximal (for the frequency $f=1 \ {\rm Hz}$) then $T_{\rm g} \approx 0^\circ {\rm C}$.

We have also performed measurements where the strain is increased slowly from zero to $\sim 100\%$.
Fig. \ref{stressStrain.eps}
shows the dependency of the stress on the strain for the soft PVC used for the suction cup A (red) and B (green).
The strain was increased from zero to its final value in about $300 \ {\rm s}$ giving a strain rate of
about $0.003 \ {\rm s^{-1}}$. The low strain modulus is about $E \approx 4.5 \ {\rm MPa}$ for both PVC compounds.

\end{document}